\documentclass[twocolumn,pdftex]{revtex4}

\usepackage[pdftex]{graphicx}
\usepackage{amsmath}

\begin{document}

\title{Radio Emission of Air Showers with Extremely High Energy Measured by the Yakutsk Radio Array}


\author{S.P. Knurenko}
\author{Z.E. Petrov}
\author{I.S. Petrov$^{*}$}

\affiliation{Yu. G. Shafer Institute of Cosmophysical Research and Aeronomy Siberian Branch of the RAoS, Russia}

\email{igor.petrov@ikfia.ysn.ru}

\begin{abstract}
  The Yakutsk Array is designed to study cosmic rays at energy 10$^{15}$-10$^{20}$ eV. It consists several independent arrays that register charged particles, muons with energy E$\geq$1 GeV, Cherenkov light and radio emission. The paper presents a technical description of the Yakutsk Radio Array and some preliminary results obtained from measurements of radio emission at 30-35 MHz frequency induced by air shower particles with energy $\varepsilon$ $\geq$ 1$\cdot$10$^{17}$ eV. The data obtained at the Yakutsk array in 1986-1989 (first set of measurements) and 2009-2014 (new set of measurements).
  Based on the obtained results we determined:

  Lateral distribution function (LDF) of air showers radio emission with energy $\geq$ 10$^{17}$ eV.
  Radio emission amplitude empirical connection with air shower energy. Determination of depth of maximum by the ratio of amplitude at different distances from the shower axis.
  For the first time, at the Yakutsk array, radio emission from the air shower with energy $>$ 10$^{19}$ eV was registered including the shower with the highest energy ever registered at the Yakutsk array with energy $\sim$2 $\cdot$ 10$^{20}$ eV.
\end{abstract}

\keywords{Radio emission, Air showers, Ultra-high energy cosmic rays, Yakutsk array}

\maketitle

\section{Introduction}

The method of registration of radio emission of ultrahigh energy particles is based on the motion of charged particles in the geomagnetic field \cite{Kahn2892061966, Falcke194772003} and Askaryan effect \cite{Askaryan1961616}. Apparently, both generation mechanisms are effective in passing air shower particles through the atmosphere. Their contribution to the generation of radio emission depends on the conditions of the shower development in the atmosphere: height of the shower maximum, zenith angle of the incoming shower and energy. Full research of radio emission mechanisms is only available on the arrays with a hybrid registration system of extensive air showers (EAS) particles: electrons, muons, Cherenkov light, ionization and radio emission.

 Influence from both mechanisms affects the symmetry of the lateral distribution of air showers radio emission as shown from the experiments. This is especially noticeable at small distances from the shower axis, where the radio emission intensity decreases significantly.

 In the following years after this discovery, there have been many experimental studies of radio emission from air showers \cite{Jelley1965327, Allan1971171}, including the Yakutsk array \cite{Artamonov1987109}. Short reviews of the air shower radio emission work can be found in \cite{ Tsarev2004149, Filonenko2001439, Huege62012016,Schroder931682017}. In paper \cite{Tsarev2004149} pointed out the possibility of registration of extensive air showers (EAS) with energies above 10$^{19}$ eV, employing radio equipment placed on the surface of the Earth and registering the radio emission by satellites on the Earth orbit. Surface arrays require a huge area of 3-5 thousand square kilometers for the registration of showers with such energies. In addition, it requires relatively quiet in terms of radio interference place in the urbanized society that is difficult to find. At the same time, satellite based arrays would allow a large solid angle which covers bigger areas and detects a larger number of air showers with highest energies. Thus, the problem of statistics of such showers would have been solved, and the spectrum of cosmic rays would be studied at energies up to 10$^{21}$ eV. But before one put this idea into practice, we need to ensure the effectiveness of this method of registration of showers with ultra-high energies. For these purposes would most be suited the currently existing large ground arrays where exist a corresponding infrastructure which can be used for registration of radio emission. Experiments on radio radiation from the EAS were actively carried out in 60-70 years of the last century. For example, the array of the Moscow State University in the 70s registered air shower radio emission at energies 10$^{16}$-10$^{17}$ eV \cite{Vernov1966157, Atrashkevich1978712}. Later, in 1986-1989, at the Yakutsk array were carried out measurements of radio emission in the energy range above 10$^{17}$ eV \cite{Artamonov1987109, Artamonov19884748, Artamonov1990210}.

In recent years, interest in the air shower radio emission, as an independent method to study the physics of the EAS has grown significantly, and for registration of radio emission were built arrays of significant size \cite{Schellart2013560, Fuchs201293}. This method makes it possible not only to evaluate the energy but also to reconstruct the longitudinal shower development, namely, the depth of maximum X$_{max}$ \cite{Apel201490, Knurenko2015632}. This is especially important for huge arrays where the uncertainty in the estimation of shower energy with different methods of detecting air showers reaches about (20-40)$\%$. For example, Auger and Telescope Array difference is 20$\%$ and the cause of differences still remains unknown \cite{Matthews20131218}. Thus, the radio emission, in conjunction with other methods could be employed for intercalibration of huge arrays \cite{Schroder010522016, Hiller7631792016, Aab93120052016}.

This paper presents radio emission of EAS with ultra-high energies data obtained by Yakutsk array in 1986-1989 and 2009-2014 years.

The paper structured as follows. In the section ~\ref{yakutsk_sec_YakutskRA} the Yakutsk array is described: frequency choice for registration, equipment, software for registration and analysis. Methodological issues like background noises at the Yakutsk array region is discussed in section ~\ref{yakutsk_sec_meth_issues}. In section ~\ref{yakutsk_sec_results} the results of registration of air showers with energy E = 10$^{17}$-10$^{18}$ eV and E$\geq$10$^{19}$ eV are presented, also energy estimation, determination of depth of maximum development and mass composition of CR. Conclusion is discussed in section ~\ref{yakutsk_sec_conclusion}.

\section{Yakutsk Radio Array}
\label{yakutsk_sec_YakutskRA}
\subsection{First Stage}
\label{yakutsk_sec_First_stage}

\begin{figure}[h]
\center{\includegraphics[width=0.5\linewidth]{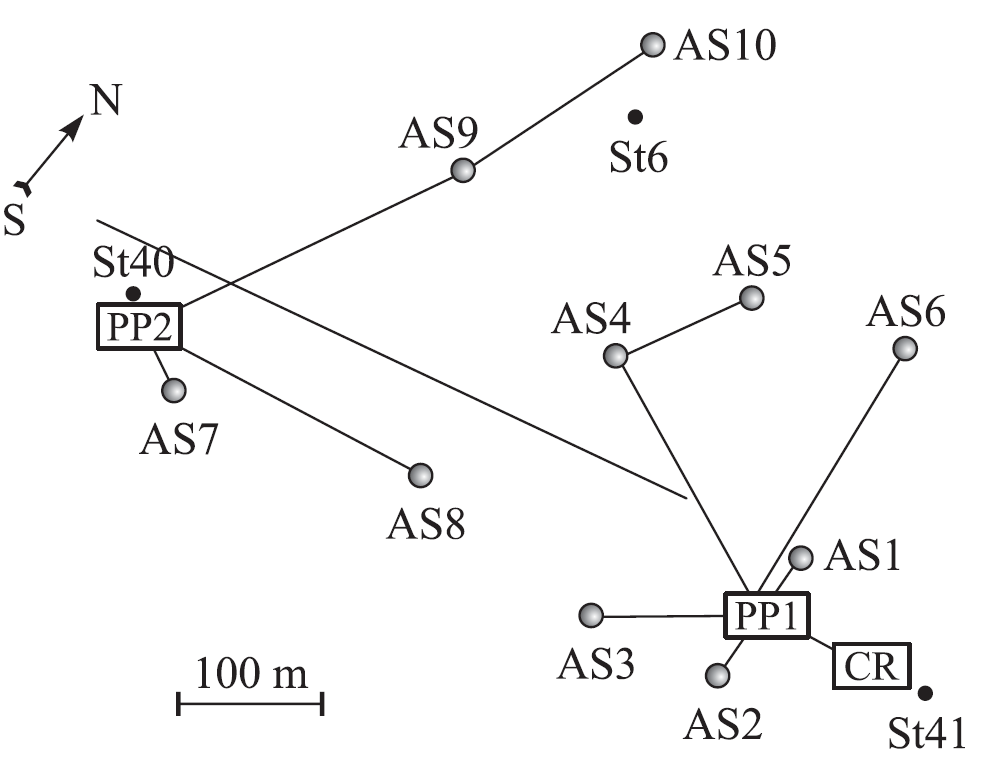}}
\caption{The arrangement of the radio antennas, in 1986 - 1989. AS - antenna station; PP - peripheral (intermediate) collection point for air shower data; CR - central registration point; ST – station with scintillation detectors of Yakutsk array}
\label{Fig:image1}
\end{figure}

In the mid 80-es of the last century, the Radio Array with registration bandwidth 30-40 MHz was designed as an extension of main Yakutsk particle array \cite{Artamonov1987109}.The setup consisted of two parts: analog and digital. The analog part comprises the reception, amplification of the radio signal, matching circuits of the output signals by the level and frequency with the parameters of the digital recorder. The digital part of the array converts input analog signals into digital code and writes information of the radio noise state and the signal from the shower to a buffer RAM. Then the information about the noise field and a radio pulse from EAS were copied to the computer hard disc drive (HDD). Air shower radio emission is registered by 20 receiving antennas, which are installed on 10 pillars as shown in Fig. \ref{Fig:image1}.

The distance between antenna pillars was 50, 100, 200, 300 and 500 meters, covering an area of roughly 0.35 km$^{2}$. One pillar consists two independent half-wave dipoles with orientation E-W and N-S. Antennas installed at $\lambda$ / 4 above ground, thus ensuring a maximum of the radiation pattern of the emission coming from the top.

Fig. 1 shows the location of antennas. At the lower part of the antenna pillar special container is located. In order to enhance the radio signals, unified broadband receivers with direct amplification at the bandwidth of 30-35 MHz were applied. Suppression of the gain at frequency 29 MHz $\geq$ 40 dB, at frequencies less than 28 MHz and more than 36 $\geq$ 60 dB.

Constructively, the receivers are designed as two blocks. The first block is located under the antenna, the block consists a low noise amplifier with a gain of Ku $\cong$  40 dB and output matching with the cable. In the second block, the final amplifier with gain Ku $\cong$ 40 dB was placed. To match with a bandwidth of the ADC at the output of the amplifier amplitude detectors are used. To improve detection of linearity powerful FET (Field Effect Transistor) type KP901 and KP902 were used.

All recording equipment: power amplifiers, detectors and ADC were placed in the two warm cabins, because of the extremely low winter temperatures (-40 $^\circ$ C). Also, the cabins contain calibration generators and high-frequency switches.

\paragraph{Calibration} During the calibration process, the input of the antenna amplifier is disconnected from the antenna and is connected to the output of the calibration generator via coaxial cable. Calibration is performed automatically without operator intervention at specified intervals of time. For this purpose, the remote-controlled generators G4-151 and RF switches on the relay of REV-15, which is controlled by a central computer, were used. To improve the accuracy of timing synchronization of additional ADC of crystal oscillators has been introduced.

\paragraph{ADC} In the first stage of the experiment, ADC F-4226 with the following parameters were used: sampling frequency - 20 MHz, conversion time - 50 ns, accuracy -8 bits (256 amplitude points), RAM-1024 word capacity (51 ms). Continuous operation of the converter allows one to store information in the memory of the radio pulses before receiving the ADC trigger signal input from the scintillation detectors of the Yakutsk array. The 9th bit of data word is a sign of the data are in the RAM to run.

Additional synchronization with EAS provided by a separate channel of signal detection for time synchronization at a frequency of 207 MHz with an accuracy of 100 ns, using the same type of ADC.

\subsection{Current State of Radio Array}
\label{yakutsk_sec_current_state}
\subsubsection{Selection of Optimal Frequency for Air Showers Radio Emission Registration}
\label {yakutsk_sec_freq}
In 2009,  for an optimal frequency choice, the background frequency spectrum from 1 to 100 MHz was analyzed \cite{Kozlov2012215}, according to work \cite{Ellingson5532007}. We used digital spectrum analyzer ASA-2332. At frequencies, up to 20 MHz due to the presence of large natural radio noise (primarily storm origin), it is not possible to distinguish air showers pulses with sufficient efficiency. Therefore, it is reasonable to select frequencies above 20 MHz, since ionosphere noises amplitude decreases dramatically in the transition to high frequencies and is about (0.5-1)$\mu$V$\cdot$m$^{-1}$$\cdot$MHz$^{-1}$ at the frequency $\sim$20 MHz. Over this frequency range, the amplitude of galactic noises decreases much slower with the frequency than storm noises. At 32 MHz it is 1.0-2.0 $\mu$V$\cdot$m$^{-1}$$\cdot$MHz$^{-1}$. Thermal noise of the antenna is much smaller than the galactic noise at frequencies up to 100 MHz and has almost no influence on our measurements. Therefore, the most favorable frequency range for the measurements at the Yakutsk array is 30-40 MHz, where expected the best signal-to-noise ratio because at higher frequencies the spectrum is limited by strong interfering man-made signals, e.g. broadcast television.

\begin{figure}[h]
\center{\includegraphics[width=0.6\linewidth]{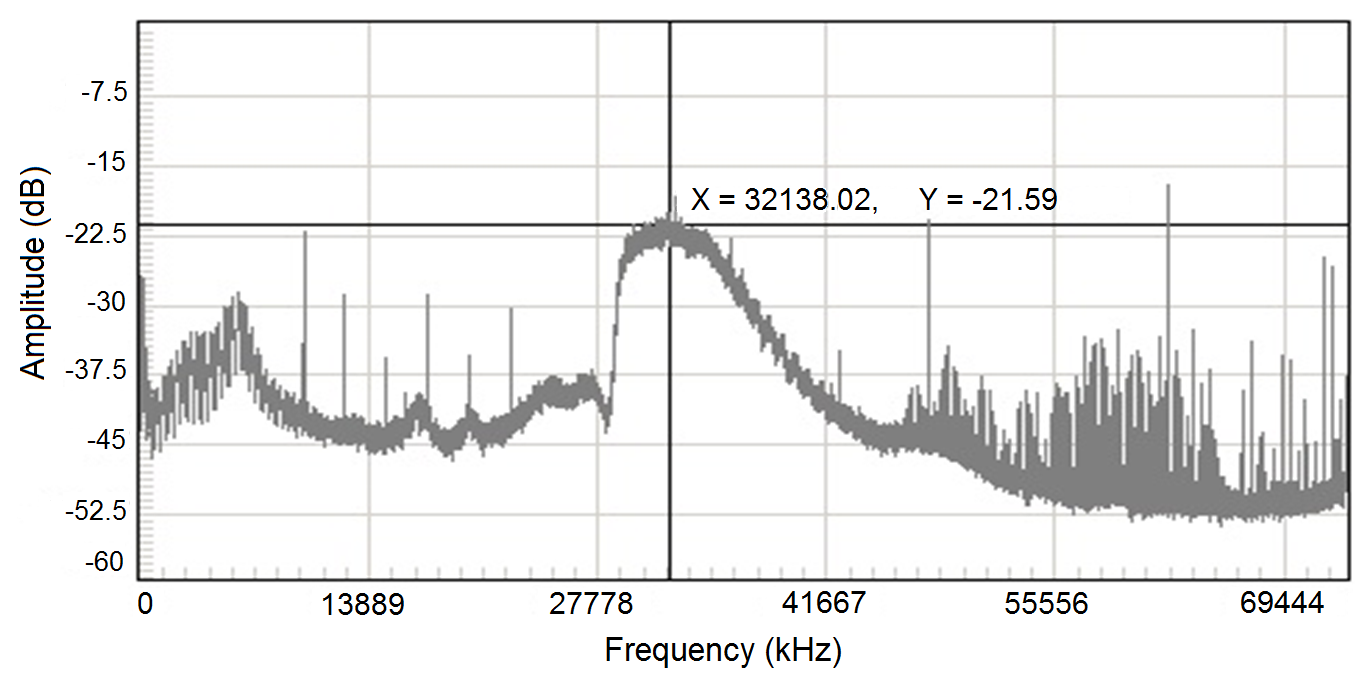}}
\caption{Frequency spectrum of radio noises at the output of the analog receiver}
\label{Fig:image2}
\end{figure}

Also, we measured background noise at the output of the analog receiver (Fig. \ref{Fig:image2}) of the Yakutsk radio array. The Fig. \ref{Fig:image2} indicates the window with a frequency of 28-42 MHz with no significant interference.

\subsubsection{Hardware and Measurement Technique}
\label{yakutsk_sec_HW_and_measu}

In early 2009, at the Yakutsk array, the radio array was reconstructed \cite{Petrov2011510}. It consists 12 crossed at 90$^\circ$ receiving antennas oriented in directions W - E (West - East), N - S (North - South),  the peripheral recording device (PRD) and data storage on a personal computer. PRD was located directly at the antenna field. The antenna field is located close to the main center of Yakutsk array and consists two independent clusters, synchronized by GPS system. The spacing between antennas was 50 m, 100 m and 500 m.  Antennas located at a distance of 50-100 m from ground stations with scintillation detectors.

Antennas that used to register high-frequency signals are shown in Fig. 3a and Fig. 3b.

\begin{figure}[h]
\begin{minipage}[h]{0.8\linewidth}
\center{\includegraphics[width=0.8\linewidth]{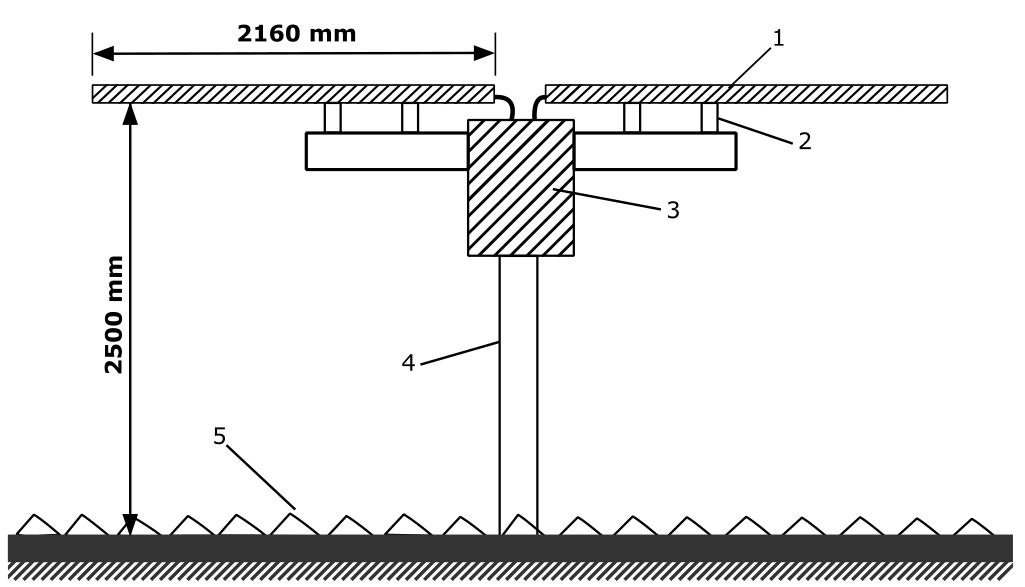} \\ a)}
\end{minipage}
\hfill
\begin{minipage}[h]{0.49\linewidth}
\center{\includegraphics[width=0.8\linewidth]{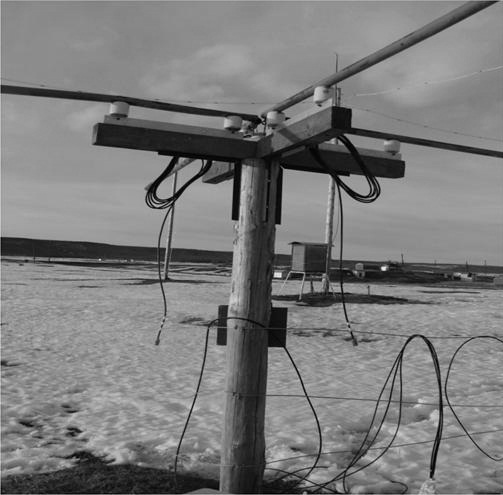} \\ b)}
\end{minipage}
\caption{Radio antenna at Yakutsk array. a) The design of receiving antenna. 1 - half-wave dipole, 2 - insulator, 3 - heat stabilized container with electronics, 4 - mounting rack, 5 - reflective screen b) Photography of receiving antenna at Yakutsk array.}
\label{Fig:image3}
\end{figure}

The bandwidth is $\pm$4 MHz, sensitivity $\sim$10$\mu$V (2 $\mu$V$\cdot$m$^{-1}$ $\cdot$ MHz$^{-1}$), the dynamic range is 50 dB. Receiving channels are based on the principle of direct signal amplification and subsequent detection. Antenna low noise amplifiers are placed into special containers (isolated metal box), connected directly to the antenna; a buried coaxial cable connects the antenna with final amplifier, filter and ADC for the digital data acquisition. The main paths are based on the cascade amplification circuit with mismatched contours. As a recorder computer, IBM PC / AT type was employed. We used fast 8-bit LA n10M8PCI as an ADC with sampling frequency 100 MHz.

\begin{figure}[h]
\center{\includegraphics[width=0.7\linewidth]{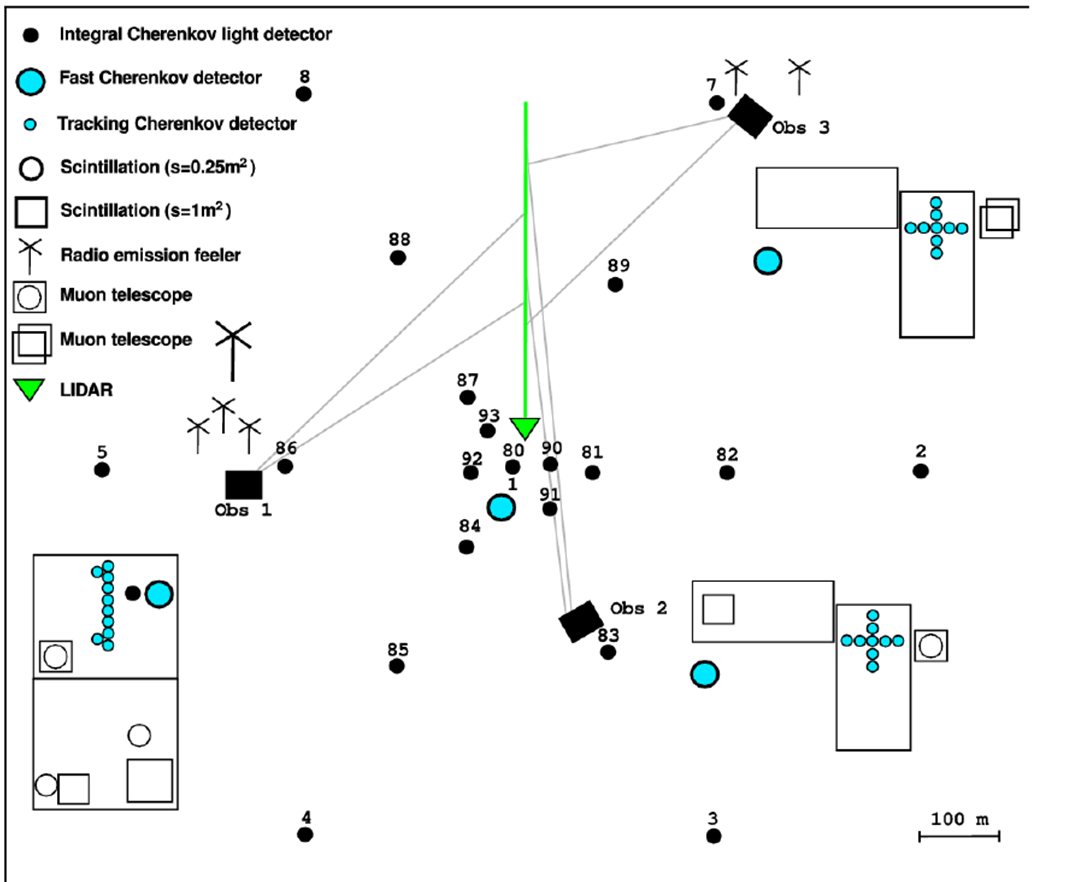}}
\caption{Arrangement of observation stations at the Small Cherenkov array}
\label{Fig:image4}
\end{figure}

Synchronization of all antennas including with a central station of Yakutsk array is provided by GPS system. It made possible to identify radio emission from air showers with a great accuracy. The location of the antenna sets is shown in Fig. \ref{Fig:image4}.

Registration of radio emission is triggered by one of two event triggers from the Yakutsk array. The first of the two possible triggers is the main Yakutsk array trigger, which registers in an area of 12 km$^{2}$ showers with energy more than 10$^{17}$ eV. The Small Cherenkov array (Fig. \ref{Fig:image4}) registers in an area of 1 km$^{2}$ showers with energy 10$^{15}$-5$\cdot$10$^{17}$ eV.

The controller registration, collection and storage of the data was performed based on an industrial computer with 19 PCI slots (can be installed 19 PCI card). The program saves a number of the antenna and the number of ADC, the date, the time, the air shower and the technical characteristics of the ADC. The sample rate of the ADC is 100 MHz, the capacity of the buffer memory is 2 MB. In addition, the ADC allows recording the prehistory (before the trigger arrival) of 25 $\mu$s and history (after the arrival of the trigger) for 15 $\mu$s.

Calibration of radio channels is performed by applying the calibration of radio pulses at the input of antenna amplifiers. signal duration was 200 ns, 32 MHz frequency filling. Calibration pulses are passed through the entire path and recorded in digital form on the hard disk along with the experimental data. 	

\subsubsection{Registration and Analysis Software. Database}

For the registration, control and calibration purposes the program (Fig. \ref{Fig:image5}) was developed. It provided a display of pulse information from each radio channel and its condition. Each frame contained information of 4 antennas colored individually and the program is able to select a single channel and view it. The program is able to load previously saved registration data, view it, use selection and group events by given criteria, create a new file with selected showers. All pulses data including calibrating pulses from every antennas are saved in separate file. Further, these data are used for mathematical processing of each shower event.

\begin{figure}[h]
\center{\includegraphics[width=0.8\linewidth]{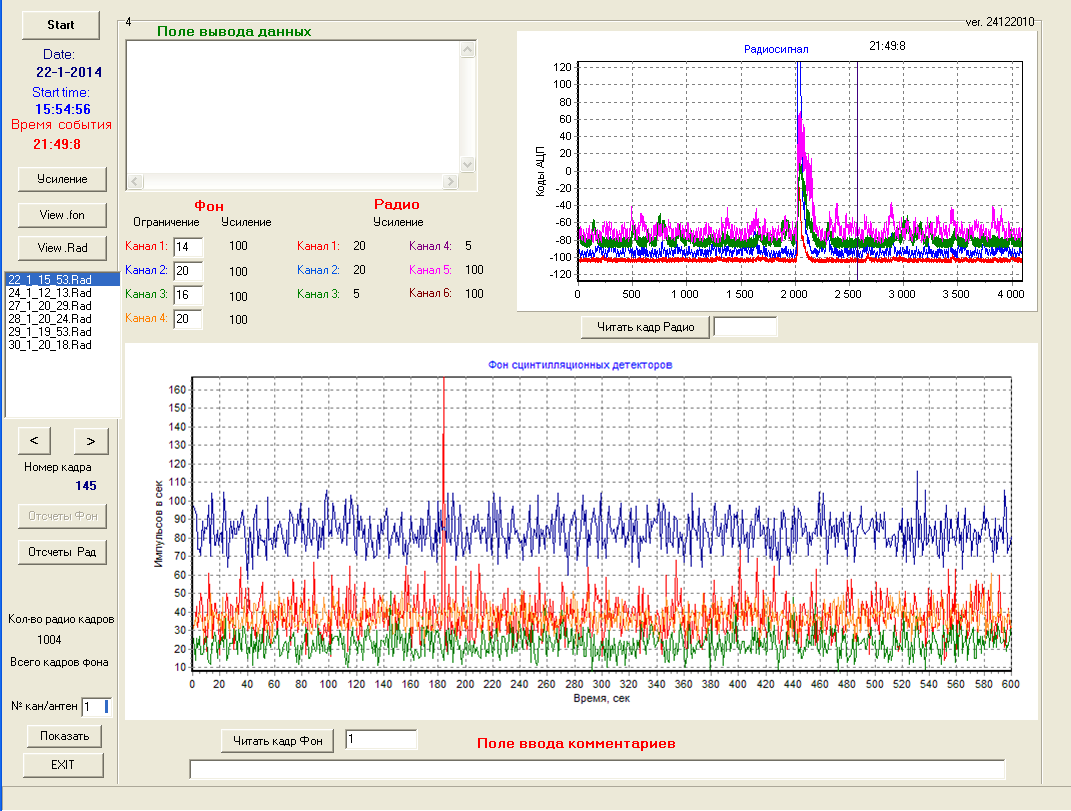}}
\caption{Program interface for noise and shower event pulses view. The showers are registered by ground scintillation detectors and radio antennas}
\label{Fig:image5}
\end{figure}

The software is able to show a background of scintillation detectors as well as radio emission background before and after registration of radio signal. In order to do this one need to select file viewing mode by clicking on the “View.Rad” button (Fig. \ref{Fig:image5}). Directly from interface panel one can set selection criteria for the registered data and create a new file with only showers that fits selection criteria. For analysis of accumulated statistics the program of preliminary selection, data accumulation and analysis was developed.

\begin{figure}[h]
\center{\includegraphics[width=0.8\linewidth]{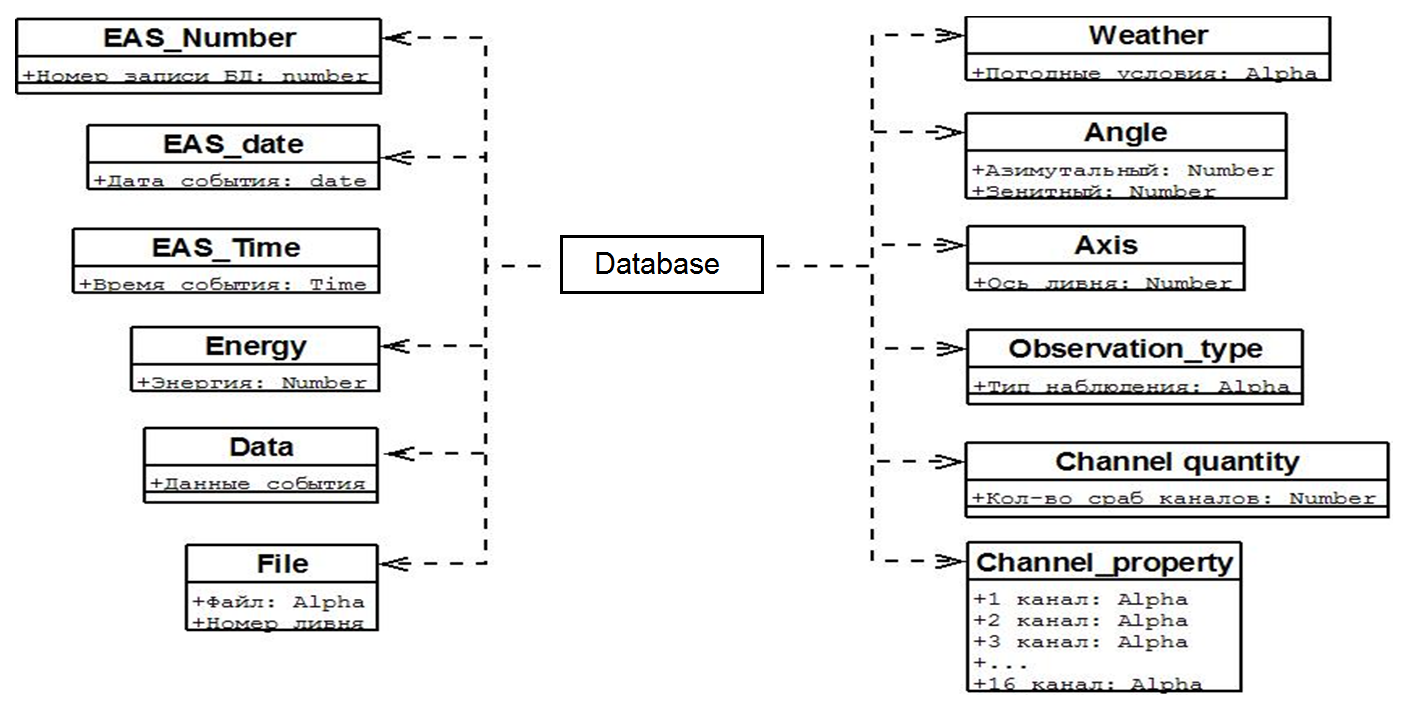}}
\caption{Flowchart of database}
\label{Fig:image6}
\end{figure}

The database is a table in Paradox 7 format (Fig. \ref{Fig:image6}). The database structure is as follows: record number, date, time, observation type, channel quantity, weather condition, event data and link to the original file, arrival angle of the shower, shower axis, primary particle energy and different time characteristics of the pulses.

Due to the small number of antennas, only 45 $\%$ of showers data contains showers with radio emission the software has additional code that enables edit function. The edit function allows one to find air shower event with no radio emission, false events, and different types of ADC failure, exclude obvious noises etc. Remaining shower events with radio emission comprised cleared database, which can be used for further analysis.

\section{Methodological Issues of Measurement of Air Showers Radio Emission}

\subsection{Radio Noise Background at Yakutsk Array Latitude}

Analysis of noise background made at Yakutsk array shown that noise background contains pulses of different nature, which can be divided by a signal duration and pulse shape. The analysis of spectrum was carried out during different seasons (autumn, winter and spring) for a week. In the total spectrum, artificial interference from high voltage electric circuits can be seen (Fig. \ref{Fig:image7}a). They are distinguished by a wide spectrum and long signal duration about milliseconds. In addition, there were signals from broadcasting stations mostly from Asia and Japanese islands. In order to exclude them we chose frequency band free of the broadcast stations (\ref{yakutsk_sec_freq}). In addition, we used trigger from the Yakutsk array to register air shower radio emission and the peak of radio emission signal should be within signal window, according to the trigger signal delay. Also, by using scintillation and Cherenkov detectors as a trigger we exclude noise signals with high amplitude registered by radio emission antennas. They feature certain set of pulses, specific appearance on the air and pulse duration. For these reasons, they are easily identified from the whole spectrum of observed pulses. Apart from man-made noises, we observed natural background noise of the Galaxy. The pulses form of this emission were close to the form of pulses from EAS.

\begin{figure}[h]
\begin{minipage}[h]{0.9\linewidth}
\center{\includegraphics[width=0.9\linewidth]{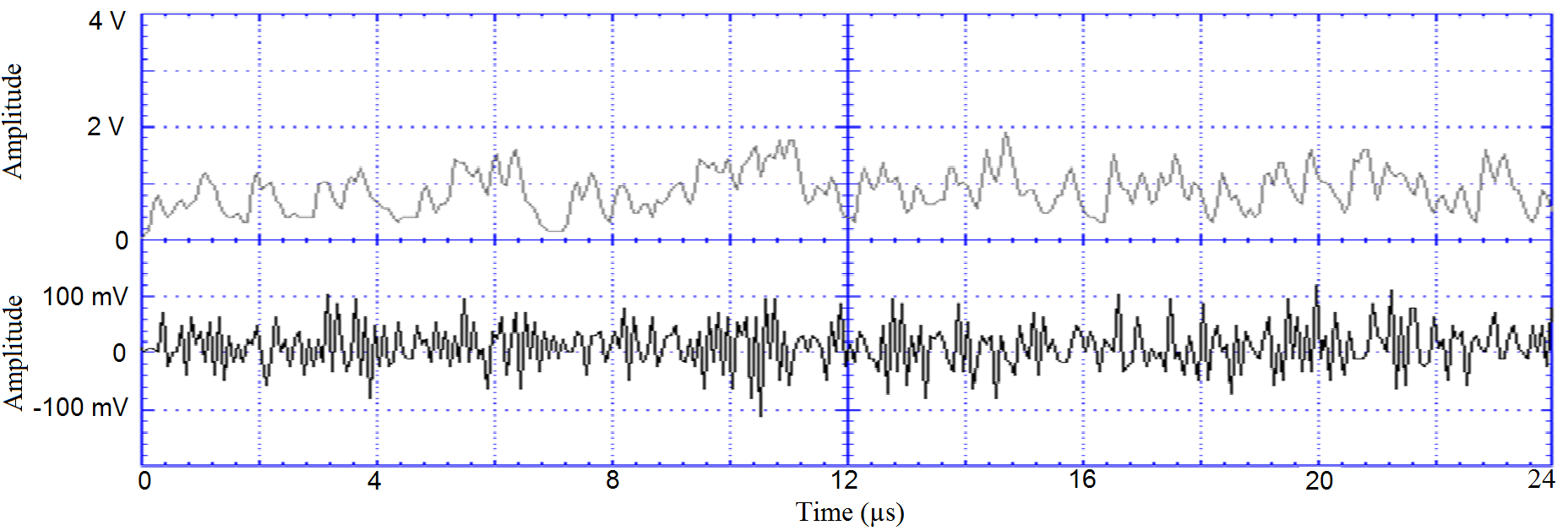} \\ a)}
\end{minipage}
\hfill
\begin{minipage}[h]{0.9\linewidth}
\center{\includegraphics[width=0.9\linewidth]{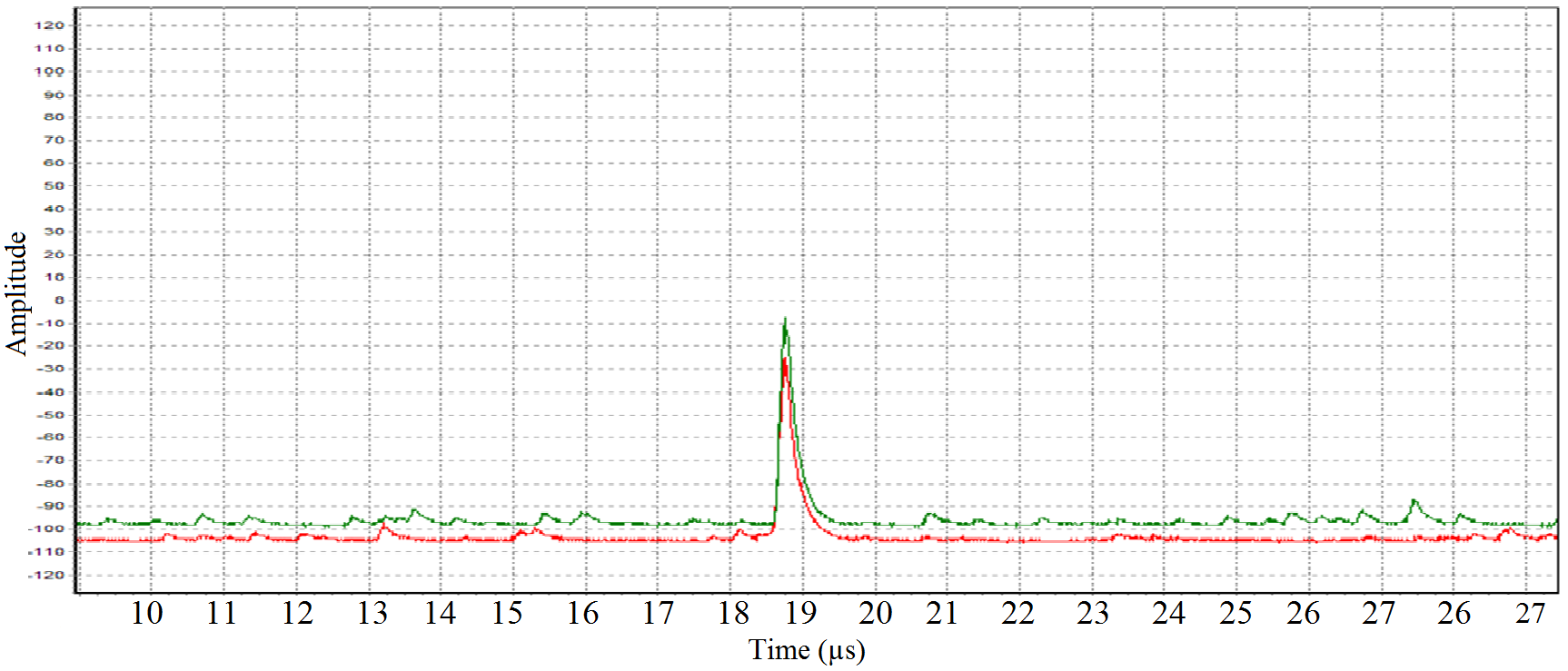} \\ b)}
\end{minipage}
\caption{Radio antenna signals at Yakutsk array. a) Background noise from Yakutsk array antenna; b) Air shower radio emission event registered by two antennas with direction W-E and N-S}
\label{Fig:image5}
\end{figure}

\subsection {Diurnal Variation}

\begin{figure}[h]
\begin{minipage}[h]{0.9\linewidth}
\center{\includegraphics[width=0.8\linewidth]{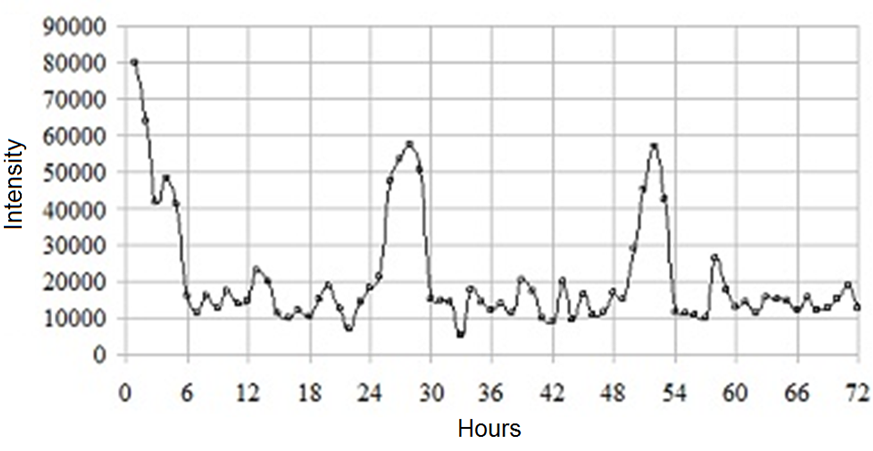} \\ a)}
\end{minipage}
\hfill
\begin{minipage}[h]{0.9\linewidth}
\center{\includegraphics[width=0.8\linewidth]{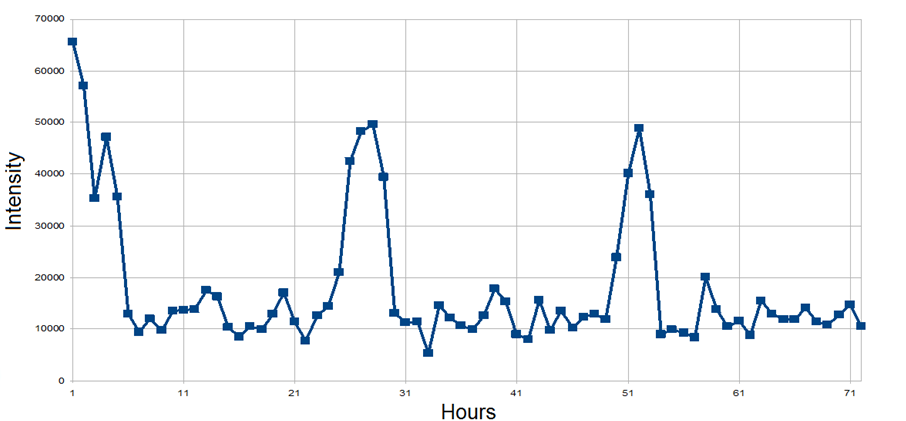} \\ b)}
\end{minipage}
\caption{Diurnal variation of radio emission. a) Antenna oriented in E-W direction. b) Antenna oriented in N-S direction}
\label{Fig:image7}
\end{figure}

In order to study the nature of the noise spectrum, we analyzed the rate of pulses for 3 days (15 - 17 October 2011). The result is shown in Fig. \ref{Fig:image7}a and \ref{Fig:image7}b.

A diurnal variation is clearly visible. This is also indicated by the difference in intensity of the spectra obtained for the day and night periods \cite{Petrov2011510}. The observed pattern in a good agreement with the result obtained in the paper \cite{Ellingson5532007}, which explains the presence of the diurnal variation of the radio noise amplitude by the non-uniform luminosity of different regions of the Galaxy. The larger amplitude at night is attributable to the plane of the Galaxy, which consists the highest number of emitting sources. Lower amplitude is typical for daytime when the antennas measures emission from the polar region of the Galaxy, which has lower luminosity than the plane of the Galaxy.

\subsection{Air Shower Radio Emission Polarization}

\begin{figure}[h]
\center{\includegraphics[width=0.7\linewidth]{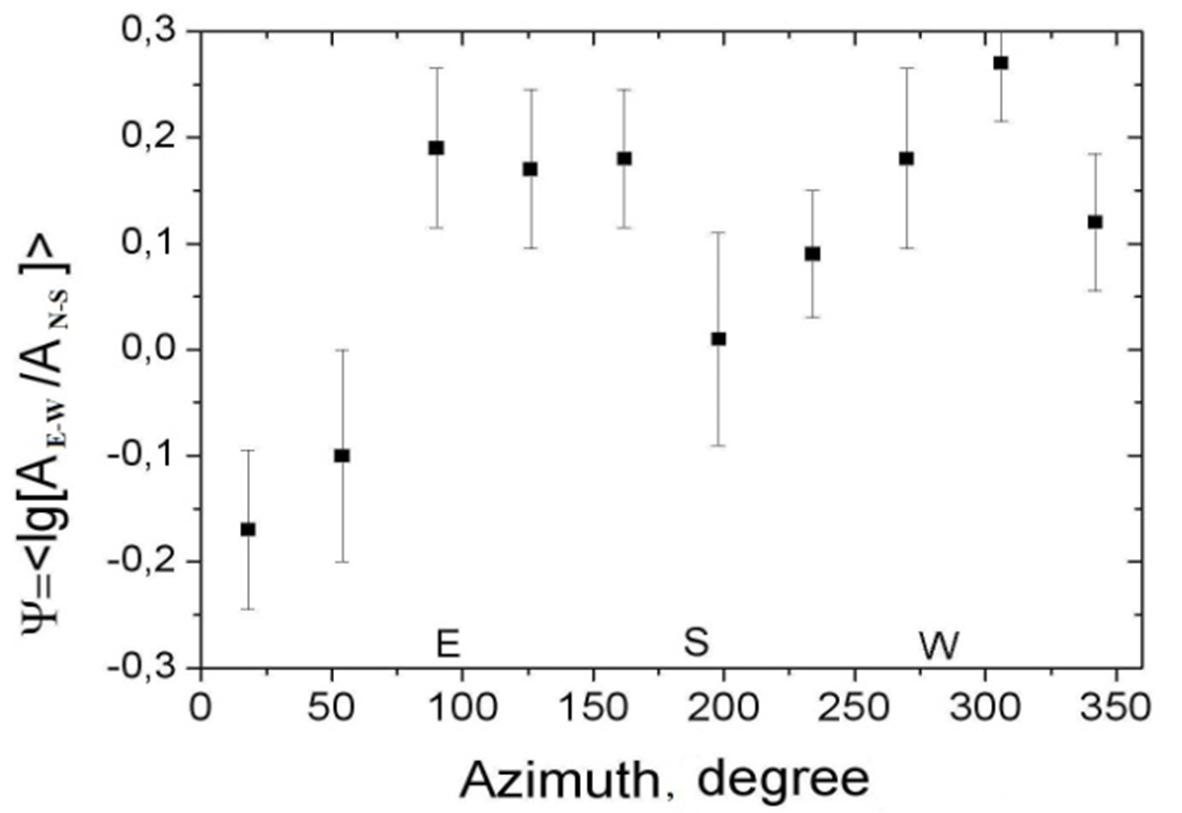}}
\caption{Dependence of antenna signal amplitude from azimuth angle}
\label{Fig:image8}
\end{figure}

In order to study polarization effect of air showers, we selected 600 events with energy $\geq$5 $\cdot$ 10$^{16}$ eV. The showers axis were located within Small Cherenkov Array perimeter of 500 m and zenith angle were $\leq$45$^\circ$.  Fig. \ref{Fig:image8} shows the dependence of amplitude ratio $\psi$ in antennas with a different orientation from azimuth angle. Preliminary analysis has shown that the amplitude of the radio signal in E-W direction is always higher than the amplitude of radio signals in N-S direction. In our case, we effectively recorded radio emission from the showers coming from the north. Thus, the radio emission from the EAS has a distinct polarization perpendicular to the direction of the geomagnetic field, which causes the polarizing effect of radio waves due to the action of the geomagnetic mechanism.

\section{Results}

\label{yakutsk_sec_results}
\subsection{Experimental Data}
\label{yakutsk_sec_experim_data}

Air shower detection effectiveness depends on spacing between stations, energy registration threshold, area of array and number of stations. In the case of the Yakutsk radio array with the area of s = 0.1 km$^{2}$ and the distances between antennas R= 50, 100 and 500 m with 12 antennas, the expected effectiveness of measurements is low. It can be seen from Table ~\ref{yakutsk_tab_total_stat} with raw data of registered air showers in the annual measurement cycle. Despite the significant observation time comparing to, for example, optical measurements, which is as low as $\sim$10$\%$ of annual measurements cycle of the Yakutsk array. A number of air showers with registered radio emission are only a small fraction. This is due to small amount of antennas and relatively small area of radio array. Besides, in showers with energy less than 10$^{17}$ eV and with distance between antenna and axis larger than 500 m radio signal is commensurable to background noise. In this case, it’s difficult to distinguish air shower radio emission among the noises.

\begin{table*}[tb!]
\centering
\caption{Air showers events statistics registered at the Yakutsk array in 2009-2016}
\label{yakutsk_tab_total_stat}       
\begin{tabular}{|p{0.1\linewidth}|p{0.1\linewidth}|p{0.1\linewidth}|p{0.1\linewidth}|p{0.1\linewidth}|p{0.1\linewidth}|p{0.1\linewidth}|p{0.1\linewidth}|}
\hline

Sea\-sons&Obser\-vation time (hours)&Total EAS events number & Per\-cen\-tage of an\-a\-lyzed events &Num\-ber of EAS with mu\-ons data & Num\-ber of EAS with Cheren\-kov light data& Cheren\-kov light ob\-ser\-va\-tion time (hours) & Num\-ber of EAS with radio emis\-sion\\\hline
2009-2010 & 6153.83 & 113138 & 87$\%$  & 60618 & 9897  & 621.78  & 822  \\\hline
2010-2011 & 6455.25 & 137830 & 89$\%$  & 56130 & 8611  & 508.39  & 1017 \\\hline
2011-2012 & 6533.94 & 155351 & 91$\%$  & 54559 & 9227  & 482.11  & 1183 \\\hline
2012-2013 & 6515.54 & 149381 & 92$\%$  & 89430 & 10219 & 591.77  & 1151 \\\hline
2013-2014 & 6446.44 & 147589 & 91$\%$  & 72110 & 7164  & 396.00  & 1123 \\\hline
2014-2015 & 6365.05 & 140101 & 72$\%$  & 82392 & 7838  & 429.34  & 840  \\\hline
2015-2016 & 5671.43 & 127490 & 81.6$\%$& 62599 & 4819  & 314.60  & 867  \\\hline
\end{tabular}
\end{table*}

A total number of air shower events with radio emission registered by radio array 2009-2016 for 44141 hours is 7003 air showers (Table ~\ref{yakutsk_tab_total_stat}).

To plot LDF, we selected 421 air showers with energy 10$^{17}$ eV and zenith angle $\theta \leq$ 45$^\circ$. In all showers, there were measurements of the total charged component, muon component and the flux of Cherenkov light. A complex measurement of the shower made it possible to establish a relationship between the magnitude of the radio signal and the different characteristics of the EAS: the energy of the electromagnetic component and the longitudinal development of the shower in the atmosphere.

\subsection{Air Showers with Energy more than 10$^{19}$ eV}
\label{yakutsk_energy_more_19}

\begin{figure}[h]
\center{\includegraphics[width=0.4\linewidth]{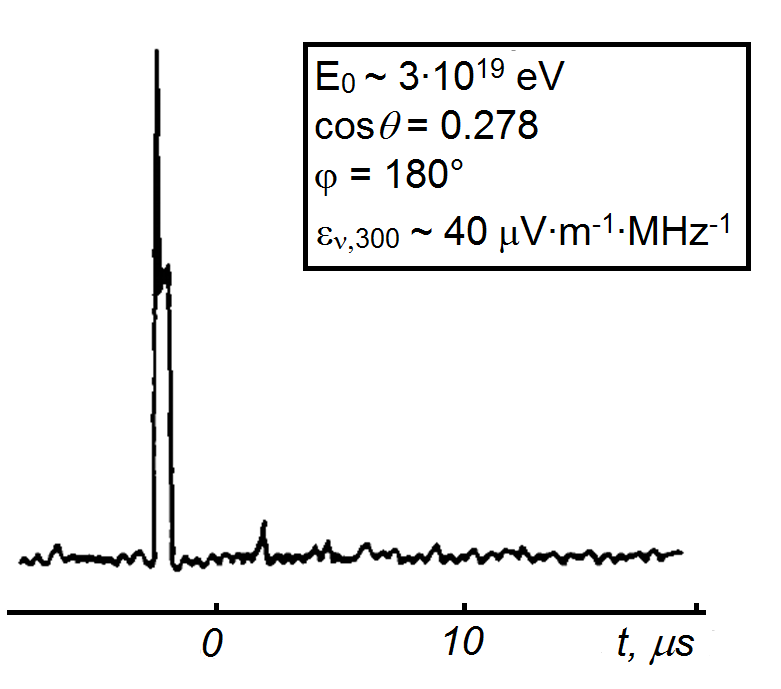}}
\caption{The shape of radio pulse from EAS with energy E$_{0}$ $\sim$ 3$\cdot$10$^{19}$ eV with zenith angle  $\geq$ 73.9$^\circ$.}
\label{Fig:image10}
\end{figure}

\begin{figure}[h]
\center{\includegraphics[width=0.6\linewidth]{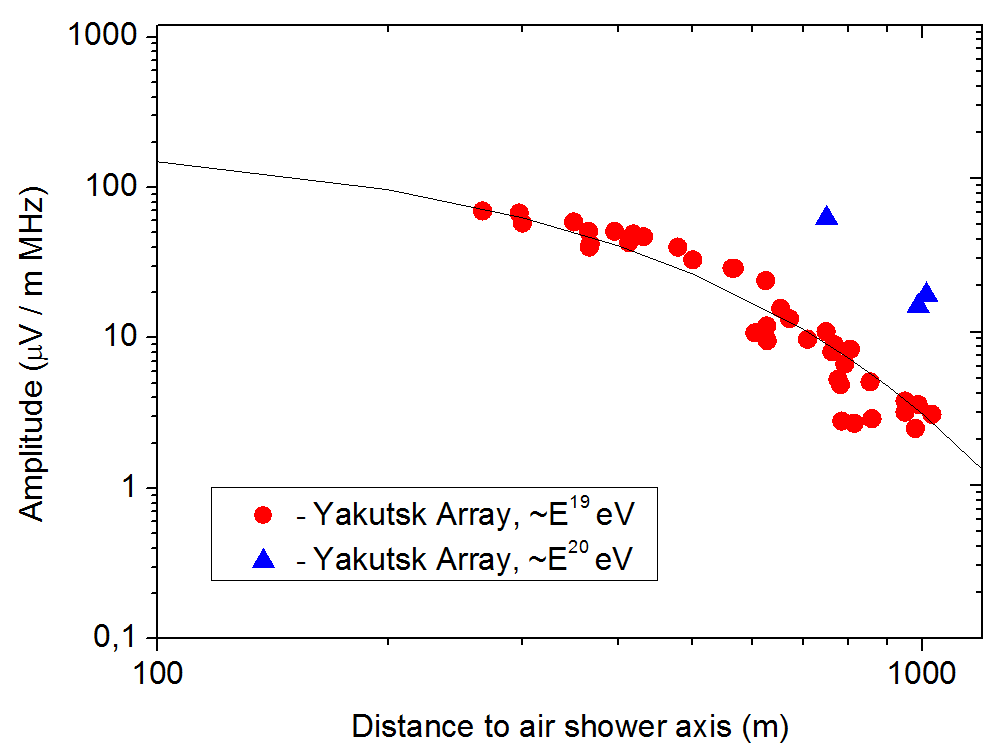}}
\caption{Lateral distribution function of air showers with energy E $\geq$ 10$^{19}$ eV. Dots are normalized to mean energy $<$E$_{0}$$>$ = 1.54$\cdot$10$^{19}$ eV and zenith angle $<\theta>$ = 43.1$^\circ$. Logarithmic scale}
\label{Fig:image11}
\end{figure}

Fig. \ref{Fig:image10} shows radio pulse of inclined air shower with energy $\sim$3$\cdot$10$^{19}$ eV \cite{Artamonov1990210}. In Table ~\ref{yakutsk_tab_stat_energy_1019} most of the showers have energy from 10$^{19}$ to 3.5$\cdot$10$^{19}$ eV, and two showers with energy $\geq$ 10$^{20}$ eV \cite{Knurenko20161045297}. Dots are normalized to mean energy $<$E$_{0}$$>$ = 1.54$\cdot$10$^{19}$ eV, to mean zenith angle $<$$\theta$$>$ = 43.1$^{\circ}$ according to eq.~\ref{yakutsk_eq_dep_theta}, eq.~\ref{yakutsk_eq4} and in logarithmic scale. The approximation is given by eq.\ref{yakutsk_eq1}. In Fig. \ref{Fig:image11} there are showers (triangles) that significantly higher than other dots. These are signal registered in two showers with maximum energy. We did not normalize these air showers to emphasize the special status of these points because they belong to the showers with energy E$_{0}$ $\geq$ 10$^{20}$ eV and their amplitude greater than in other showers.

We see that dots indicates rapid attenuation of the radio signal and shows LDF dependence on the distance of antennas from air shower axis.

Table ~\ref{yakutsk_tab_stat_energy_1019} shows air showers with the highest energy, registered at Yakutsk array in 1986-1989, 2009-2014 years.

\begin{table}
\centering
\caption{List of air showers with energy 10$^{19}$ eV registered by Yakutsk array antennas. Date - is a date of shower registration, $\theta$ - zenith angle (degree), $\psi$ - azimuth angle (degree), E$_{0}$ – energy of primary particle (eV), A$_{\nu}$ – radio emission amplitude ($\mu$V$\cdot$m$^{-1}$$\cdot$MHz$^{-1}$), R - distance from air shower axis to antenna (m)}
\label{yakutsk_tab_stat_energy_1019}       
\begin{tabular}{|l|l|l|l|l|l|}
\hline
date     & $\theta$,deg & $\psi$, deg  & E$_{0}$, eV           & A$_{\nu}$     & R, m \\\hline
16.11.86 & 74           & 180          & 3.1$\cdot$10$^{19}$   & 58.0          & 300  \\\hline
16.12.87 & 71           & 178          & 3$\cdot$10$^{19}$     & 40            & 367  \\\hline
21.02.88 & 70           & 210          & 10$^{19}$             & 3.1, 3.8      & 1030, 950 \\\hline
09.03.88 & 36           & 125          & 9$\cdot$10$^{18}$     & 6.2           & 792 \\\hline
07.05.89 & 59           & 168          & 2$\cdot$10$^{20}$     & 62.5          & 750 \\\hline
10.03.11 & 51           & 239          & 1.1$\cdot$10$^{19}$   & 89, 43, 5.8   & 350, 413, 604 \\\hline
16.05.11 & 69           & 99           & 1.6$\cdot$10$^{19}$   & 33, 29, 40    & 501, 564, 479 \\\hline
31.12.11 & 15           & 165          & 1.1$\cdot$10$^{19}$   & 1.2, 1.0, 2.9 & 950, 980, 860 \\\hline
12.04.12 & 8            & 222          & 1.3$\cdot$10$^{19}$   & 4.1, 2.8, 6.0 & 762, 785, 626 \\\hline
04.05.13 & 46           & 295          & 1.1$\cdot$10$^{19}$   & 5.3, 6.0, 12  & 776, 768, 368 \\\hline
12.12.13 & 15           & 297          & 1.2$\cdot$10$^{19}$   & 5.1, 8.4, 3.6 & 855, 806, 988 \\\hline
03.10.13 & 21           & 21           & 1.1$\cdot$10$^{19}$   & 9.1, 11, 2.7  & 419, 396, 815 \\\hline
22.03.13 & 46           & 4            & 1.8$\cdot$10$^{19}$   & 41, 48, 78    & 418, 432, 366 \\\hline
02.01.14 & 48           & 207          & 1.0$\cdot$10$^{20}$   & 16.3, 19.4    & 1013, 988     \\\hline
22.01.14 & 47           & 189          & 1.1$\cdot$10$^{19}$   & 107.6, 119.6  & 297, 266     \\\hline
05.02.14 & 26           & 343          & 3.5$\cdot$10$^{19}$   & 3.4, 5.6      & 671, 627     \\\hline
02.03.14 & 30           & 217          & 1.2$\cdot$10$^{19}$   & 4.9, 6.0, 7.8 & 782, 749, 708\\\hline
\end{tabular}
\end{table}

\subsection{LDF of Air Showers Radio Emission}
\label{yakutsk_sec_LDF}

\begin{figure}[h]
\center{\includegraphics[width=0.6\linewidth]{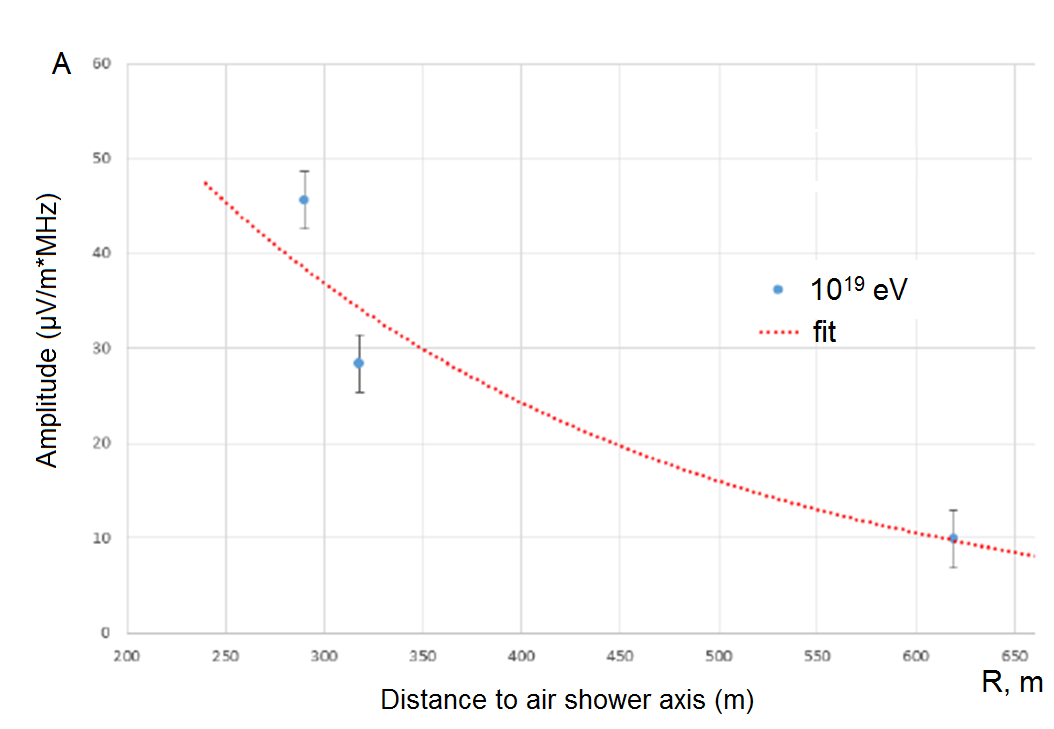}}
\caption{LDF of individual shower with energy 10$^{19}$ eV}
\label{Fig:image12}
\end{figure}

In Fig.\ref{Fig:image12} shown the signals registered by antennas at different distances from the shower axis. The vertical lines marked errors related to the absolute calibration of the radio signal. As one can see that radio emission amplitudes are significantly attenuated with distance. This picture of the formation of the spatial distribution of radio emission indicated its dependence from the longitudinal development of air showers. This is confirmed by calculations and can be used to recover the depth of maximum development X$_{max}$\cite{Huege200830,Kalmykov2013409}. In order to do that it’s enough to measure an amplitude of radio emission at different distances from air shower axis and use their ratio to estimate X$_{max}$ as shown from model calculations and experiments at LOPES, Tunka-Rex and Yakutsk \cite{Apel201285, Schroder010522016, Knurenko2015410412}.

\begin{figure}[h]
\center{\includegraphics[width=0.6\linewidth]{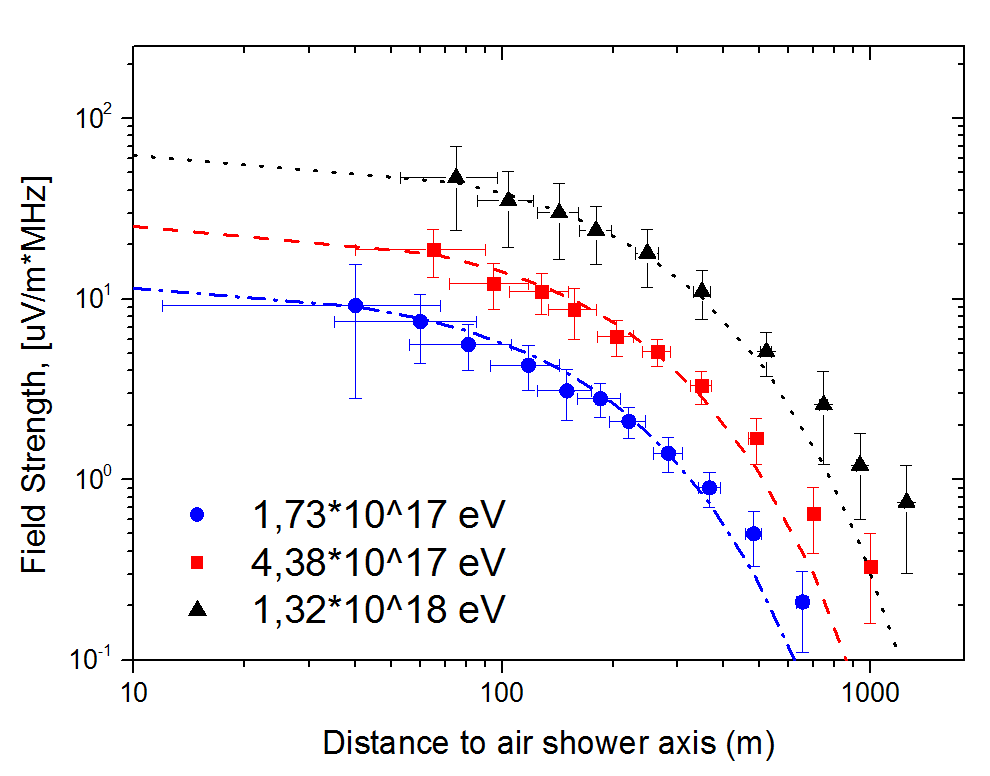}}
\caption{Average LDF of radio emission at frequency 30-35 MHz in showers with energy 1.73$\cdot$10$^{17}$ eV, 4.38$\cdot$10$^{17}$ eV and 1.32$\cdot$10$^{18}$ eV}
\label{Fig:image13}
\end{figure}

To plot average LDF, we selected showers with zenith angle $\theta$ $\leq$ 45$^\circ$- and with shower axis within the circle around the center of Yakutsk array with radius 500 m. Then, selected showers were divided into 3 groups with energy 1.73$\cdot$10$^{17}$ eV, 4.38$\cdot$10$^{17}$ eV and 1.32$\cdot$10$^{18}$ eV. The resulted set of showers were split in the evenly spaced distance of $\Delta$lgR = 0.4 m and at each interval average amplitude was calculated based on local gradient \cite{Knurenko20130054}. Additionally, we took into account noise and fluctuations of the signal at the threshold of each receiving antennas (zero points) according to \cite{Schroder2012662}. Because of this for analysis, we selected air showers with signal-to-noise ratio $\geq$5, i.e. showers within distances range of 50-500 m depending on the shower energy. For showers with energies above 3$\cdot$10$^{18}$ eV, the distance interval shifted towards higher distances R. Dependence of amplitude from distance for each energy range is shown in Fig. \ref{Fig:image13}, normalized to zenith angle $\theta$ = 35$^\circ$. The approximation curves is given by:

\begin{equation}\label{yakutsk_eq1}
  A = \varepsilon\cdot exp \left(-\frac{R}{R_0}   \right)
\end{equation}
where A - radio emission amplitude, $\varepsilon$ - the fit parameter(depends on the energy), R - the distance from shower axis to the antenna, R$_0$ - the slope parameter.

As seen from Fig. \ref{Fig:image13}, the shape of LDF is fitted well by formula (\ref{yakutsk_eq1}) only in the range of 50-350 m from the shower axis at greater distances for its approximation, we probably need another exponent.

Fig. \ref{Fig:image13} shows that slope of LDF increases with energy. Possibly, this change in the slope of LDF is related to the development of X$_{max}$ in the atmosphere. This fact we used to estimate X$_{max}$ by ratio P$_1$ = A$_1$(80)/A$_2$(200) in showers with energy E $\leq$ 3$\cdot$10$^{17}$ eV and P$_2$ = A$_1$(175)/A$_2$(725) in showers with energy E $\geq$ 3$\cdot$10$^{18}$ eV. From long-term data, we found the dependence of radio emission amplitude on zenith angle $\theta$, energy E and depth of maximum development X$_{max}$ \cite{Knurenko2015410412}. The approximation was chosen by maximum likelihood estimation:

\begin{eqnarray}\label{yakutsk_eq3}
  \varepsilon = (188.8\pm1.6)\left(\frac{E}{5\cdot10^{17}}\right)^{0.83\pm0.03}
  \cdot(1.16-\cos\alpha)\cdot\cos\theta \nonumber\\
  \cdot\exp
  \left\{
   - \frac{R}{162\pm8 + (84\pm3)
   \left[
   \frac
   {X-675}
   {100}
   \right]}
    \right\}
\end{eqnarray}
Where, $\theta$ - zenith angle, $\alpha$ - geomagnetic angle (13.8 $^\circ$ for the Yakutsk array), R - the distance to the antenna, E - the energy of the primary particle, X - the depth of the maximum development of the air shower.

From equation (\ref{yakutsk_eq3}) it follows that radio emission LDF depends on energy E and depth of maximum X$_{max}$.

\subsection{Energy Estimation}
\label{yakutsk_sec_energy_est}
\subsubsection{Energy balance method}
The method of determining the energy at the Yakutsk array is based on the expression (\ref{yakutsk_eq_energy_balance}) used in\cite{Dyakonov2521991, Knurenko83114732006}:

\begin{equation}\label{yakutsk_eq_energy_balance}
 E_0 = E_{ei} + E_{eph} + E_{\mu\nu} + E_d
\end{equation}

In the first approximation, the sum of the components presented in (\ref{yakutsk_eq_energy_balance}) will be the total energy of the primary particle. All summand of E$_0$ were reconstructed with use of integral parameters of air shower. In our case, they are determined by the following method. The energy, scattered in the atmosphere above observation level by electrons is determined by formula (\ref{yakutsk_eq_energy_electron}):

\begin{equation}\label{yakutsk_eq_energy_electron}
E_{ei} = k(x,P_\lambda)\cdot F
\end{equation}
Here F - total flux of Cherenkov light of air shower; k(x,P$_\lambda$) - fitting coefficient that takes into account the transparency of the real atmosphere and expressed through the depth of the development of the EAS measured at the array\cite{Ivanov131610012007}:

\begin{equation}\label{yakutsk_eq_energy_eph}
E_{eph} = 2.2\cdot10^6\cdot N_s(X_0)\cdot\lambda_{eff}
\end{equation}
where N$_s(X_0)$ - total number of charged particles at the sea level, $\lambda_{eff}$ - mean free path of air shower particles, which was found from correlation parameters N$_s$(X) - Q(400) at different zenith angles $\theta$ \cite{Glushkov57911993}. The energy transferred into muons with the threshold energy $\geq$ 1 GeV, estimated by the formula (\ref{yakutsk_eq_energy_muon})

\begin{equation}\label{yakutsk_eq_energy_muon}
E_mu = \varepsilon_{\mu} \cdot N_\mu
\end{equation}

Where $\varepsilon_\mu$ - average muon energy, calculated according to the measured energy spectrum of air shower muons up to Е$\mu$ = 10$^{3}$ GeV  and equal to 10.6 GeV. N$_{\mu}$ - total number of muons with threshold energy E$_{thr} \geq$ 1 GeV.

Remaining insignificant part ($\sim$5$\%$) of primary energy that is difficult to estimate experimentally we determined from calculations. Magnitude of ionization losses of muons is equal to  E$_{\mu i}$ = (0.12$\pm$0.09) E$_\mu$, energy losses on nuclear fission in the air as 0.5 GeV, ionization losses of hadron component in the atmosphere as E$_{hi}$ = (5.6 $\pm$ 2.2)$\cdot$10$^{-2}\cdot$E$_{ei}$ and the energy corresponding for neutrino component as E$_\nu$ = (0.64 $\pm$ 0.18)$\cdot$E$_{\nu}$.

The uncertainty of the method for estimating total energy of air shower in our case is 25$\%$. Further, this energy estimation is used to establish a relation of the radio emission amplitude with EAS energy \cite{Knurenko2015410412}.

\subsubsection{Radio emission correlation with air shower energy}

\begin{figure}[h]
\center{\includegraphics[width=0.5\linewidth]{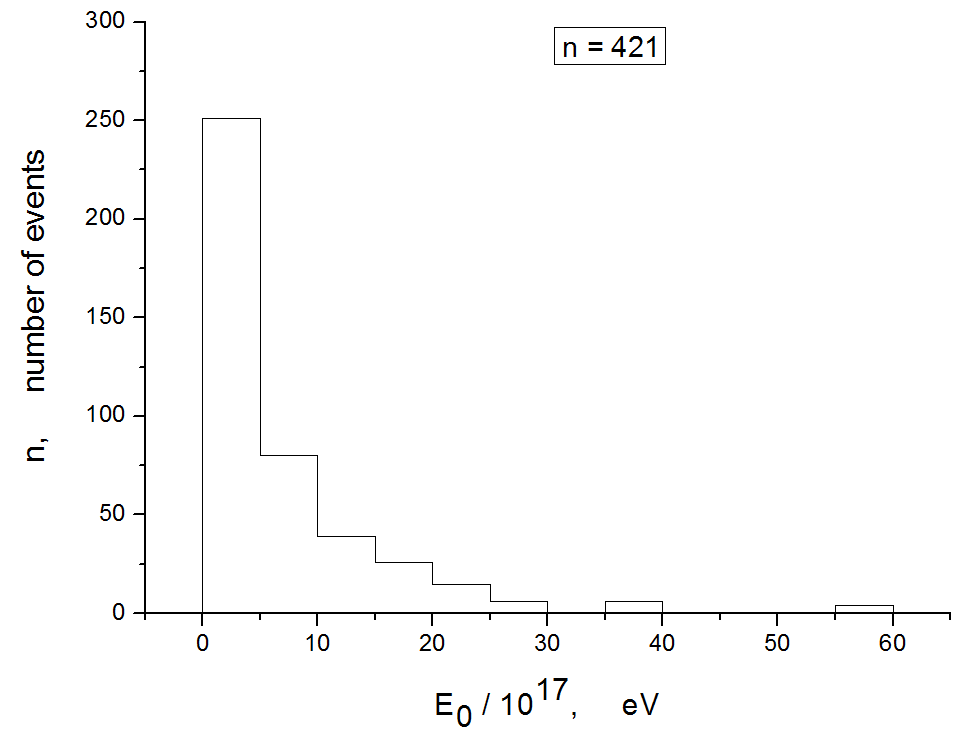}}
\caption{Air shower distribution by energy E}
\label{Fig:image13_4}
\end{figure}

\begin{figure}[h]
\center{\includegraphics[width=0.5\linewidth]{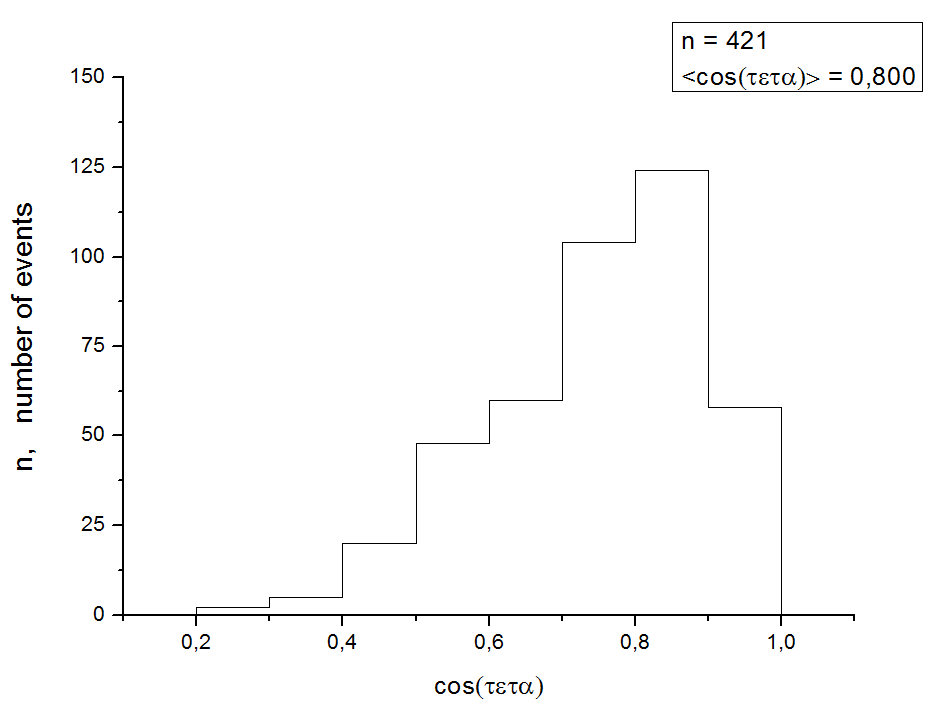}}
\caption{Air shower distribution by zenith angle (cos$\theta$)}
\label{Fig:image13_5}
\end{figure}

Air shower data, plotted against energy and zenith angle in Fig. ~\ref{Fig:image13_4}. OY axis - number of n, OX axis - air shower energy. In Fig. ~\ref{Fig:image13_5} OY axis - number of n, OX axis - zenith angle $\theta$.

\begin{figure}[h]
\center{\includegraphics[width=0.5\linewidth]{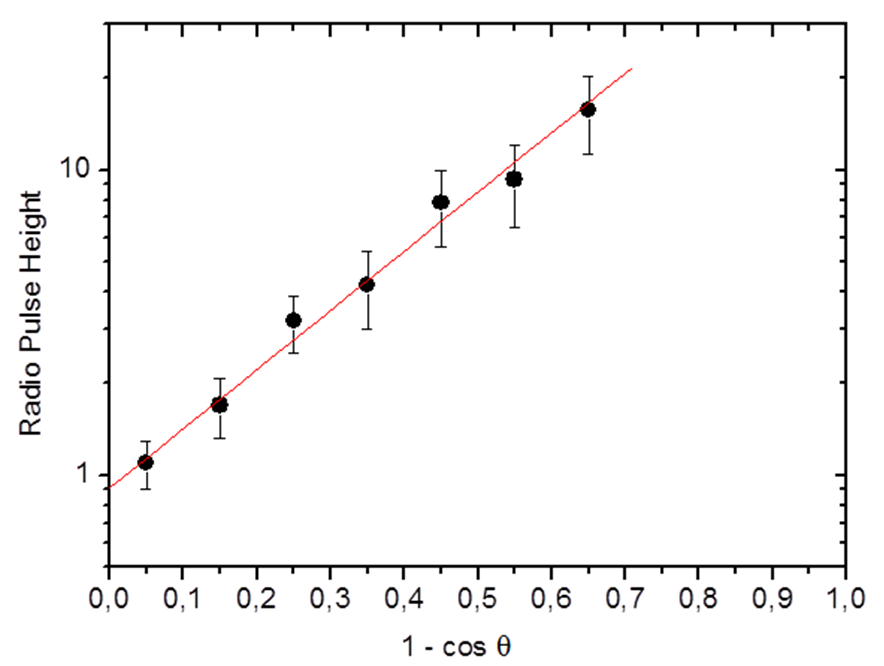}}
\caption{Dependence of radio emission amplitude on zenith angle}
\label{Fig:image13_6}
\end{figure}

Selected showers were used to plot dependence of radio emission amplitude on zenith angle $\theta$ (Fig. ~\ref{Fig:image13_6}). Obtained dependence is well described by formula:

\begin{equation}\label{yakutsk_eq_dep_theta}
\varepsilon_{EW} = (0.81\pm0.25)(1-\cos\theta)^{1.16\pm0.05}
\end{equation}

\begin{figure}[h]
\center{\includegraphics[width=0.5\linewidth]{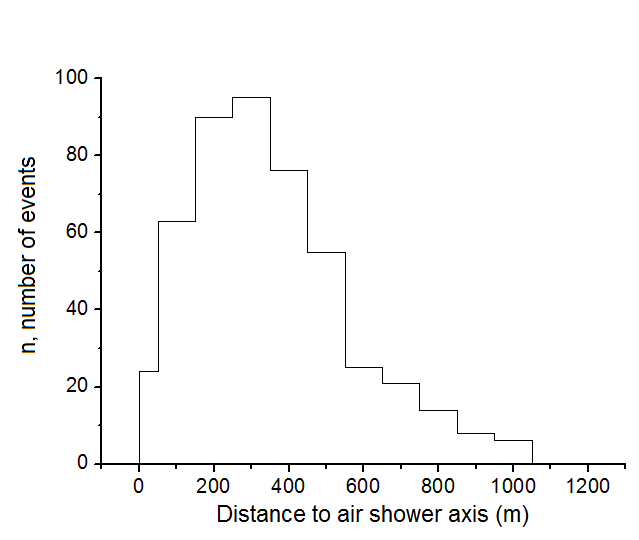}}
\caption{Distribution of the distances from antennas to air showers}
\label{Fig:image14}
\end{figure}

\begin{figure}[h]
\center{\includegraphics[width=0.5\linewidth]{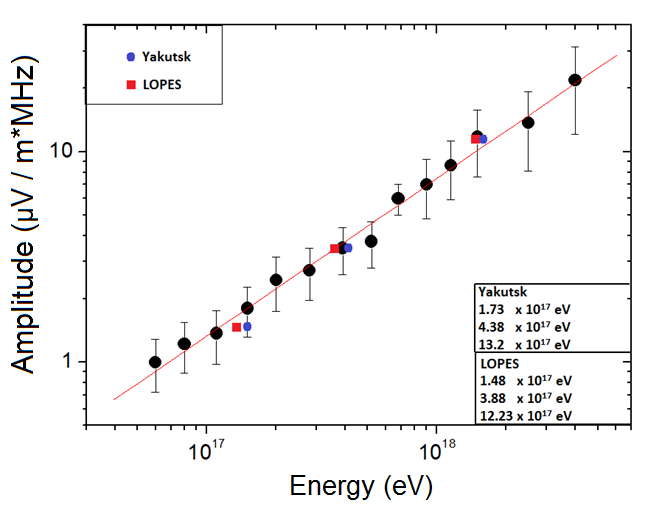}}
\caption{The dependence of the amplitude of a radio pulse from air shower energy. Data are normalized to geomagnetic angle, the average distance from the antenna to shower axis. Black dots - normalized Yakutsk data, blue dots - energy estimated by LDF data at a distance of 350 m from air shower axis, red squares - energy estimated by LOPES LDF data \cite{Horneffer20088386, Apel2010202216620}, taking into account the conditions of the Yakutsk array}
\label{Fig:image15}
\end{figure}

Fig. \ref{Fig:image14} shows the distribution of the distances from antennas to air showers. In Fig. \ref{Fig:image14} shown that the majority of antennas are in the range of 200 - 500 m. For this reason, median average distance $<$R$_{med}>$ = 350 m was taken for normalization. The amplitude of antennas at different distances from shower axis was normalized to $<$R$_{med}$$>$. Normalized radio signals were used to plot dependence of radio emission amplitude from air shower energy (Fig. \ref{Fig:image15}). Dependence of amplitude on energy is obtained from individual shower analysis in the paper  \cite{Knurenko771115592013}(eq. \ref{yakutsk_eq4}). The energy was estimated from air shower Cherenkov total light flux \cite{Knurenko83114732006}.

\begin{equation}\label{yakutsk_eq4}
  \varepsilon_{EW} = (1.3\pm0.3)\left(
  \frac{E_0}{10^{17}} eV
  \right)^{0.99\pm0.04}
\end{equation}
Here $\varepsilon_{EW}$ denotes the estimated radio amplitude and $E_0$ the energy of the primary particle.

\subsection{Estimation of Depth of Maximum Development X$_{max}$ by Radio Emission LDF}
\label{yakutsk_sec_est_depth}

Depth of maximum development of air shower X$_{max}$ at Yakutsk array determined by cascade curve reconstructed by lateral distribution of Cherenkov light using inverse problem method \cite{Tikhonov2581977} The algorithm described in details in works \cite{Dyakonov1986224226, Knurenko2001157160}. A Fredholm equation of the first kind (\ref{yakutsk_eq5}) is the basis of the algorithm, which in our case is solved by the adaptive method \cite{Kochev62711985}.

\begin{equation}\label{yakutsk_eq5}
\begin{split}
   Q_{exp} = \delta_{Q} + \int_{X_{1}}G(R,X/X_{2})\cdot N(E_0, X)\cdot K(\lambda, X)dX
\end{split}
\end{equation}
where $G(R,X/X_2)$ - the function defined by lateral-angular distribution of electrons in partial electron-photon cascade, $N(E_0, X)$ - cascade curve; $\delta_Q$ - level of "noises", depends on measurement uncertainties, statistical processing of the data, function $G(R,X/X2)$, integration etc.; $K(\lambda, X)$ - transmittance of the atmosphere; $X_1$ and $X_2$ - upper and lower limits of the atmosphere area.

As seen from eq. (\ref{yakutsk_eq5}) the method takes into account physics of air shower electron-photon component development and characteristics of the atmospheric conditions during registration of the Cherenkov radiation \cite{Dyakonov3153191999}.

Next, we had obtained the empirical relation of Cherenkov light LDF slope P(Q1/Q2) with X$_{max}$ for vertical and inclined showers \cite{Dyakonov2521991,Knurenko72512552011}.

In Fig. ~\ref{Fig:image16}a and Fig. ~\ref{Fig:image16}b correlation of LDF slope with $X_{max}$ is shown. The correlation was found by analyzing air showers registered by Cherenkov detectors and radio antennas \cite{Knurenko2015410412}. Obtained result indicates the dependence of radio emission LDF on the depth of maximum development X$_{max}$ that allows studying physics of air showers by radio emission data.

X$_{max}$ correlation with a parameter P$_1$ = A$_1$ (80) / A$_2$ (200) P$_2$ = A$_1$ (175) / A$_2$ (725) is shown in Fig. ~\ref{Fig:image16}a and Fig. ~\ref{Fig:image16}b. The solid line is simple approximation:

\begin{figure}[h]
\begin{minipage}[h]{0.7\linewidth}
\center{\includegraphics[width=0.8\linewidth]{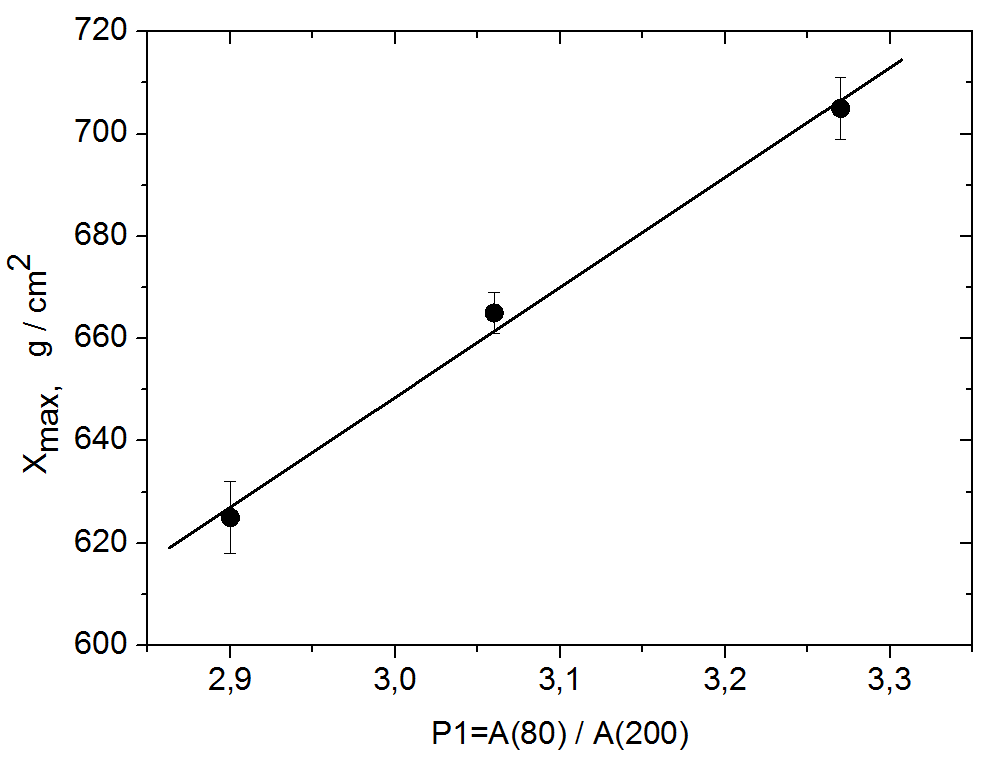} \\ a)}
\end{minipage}
\hfill
\begin{minipage}[h]{0.7\linewidth}
\center{\includegraphics[width=0.8\linewidth]{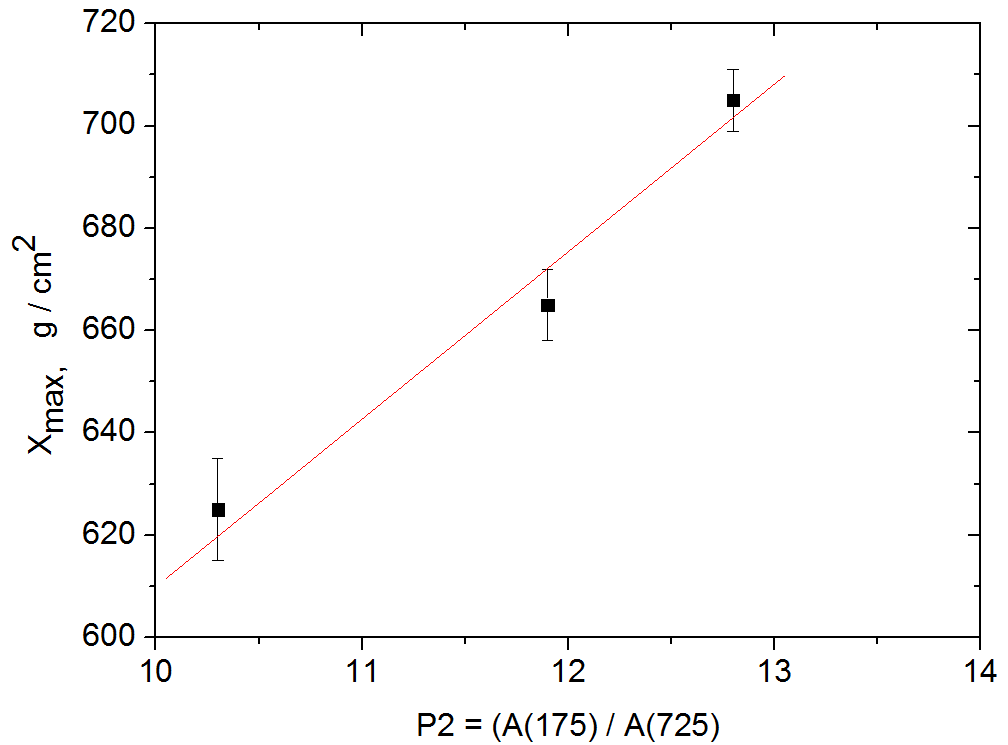} \\ b)}
\end{minipage}
\caption{X$_{max}$ correlation with the ratio of radio emission amplitudes at different distances. a) at distances 80 m and 200 m b) at distances 175 m and 725 m}
\label{Fig:image16}
\end{figure}

\begin{equation}\label{yakutsk_eq6}
\begin{split}
   X_{max} = (655 \pm 10) + (100 \pm 5)\left( (P_1 - 3.01)/0.46 \right), \\
   X_{max} = (660 \pm 15) + (100 \pm 5)\left( (P_2 - 11.5)/3 \right).
\end{split}
\end{equation}

where $P_1$ is the ratio of amplitudes at 80 m $A_1$(80) and 200 m $A_2$(200), $P_2$ is the ratio of amplitudes at 175 m $A_1(175)$ and 725 m $A_2 (725)$, $A_1$ and $A_2$ are determined by the eq.\ref{yakutsk_eq1}.

To estimate the depth of maximum X$_{max}$ in individual showers we used formula (\ref{yakutsk_eq6}). The formulas have the uncertainty of 10-25 g/cm$^{2}$  in the range of 600-790 g/cm$^{2}$ as shown by model calculation. Near sea level accuracy of X$_{max}$ estimation decreases. In general, the accuracy of measurement of the amplitude of the radio signal and EAS parameters reconstructed by radio emission depends on the error of the electronics (ADC, preamplifier and postamplifier), mathematical processing methods and a systematic uncertainty of energy estimation.

\begin{figure}[h]
\center{\includegraphics[width=0.7\linewidth]{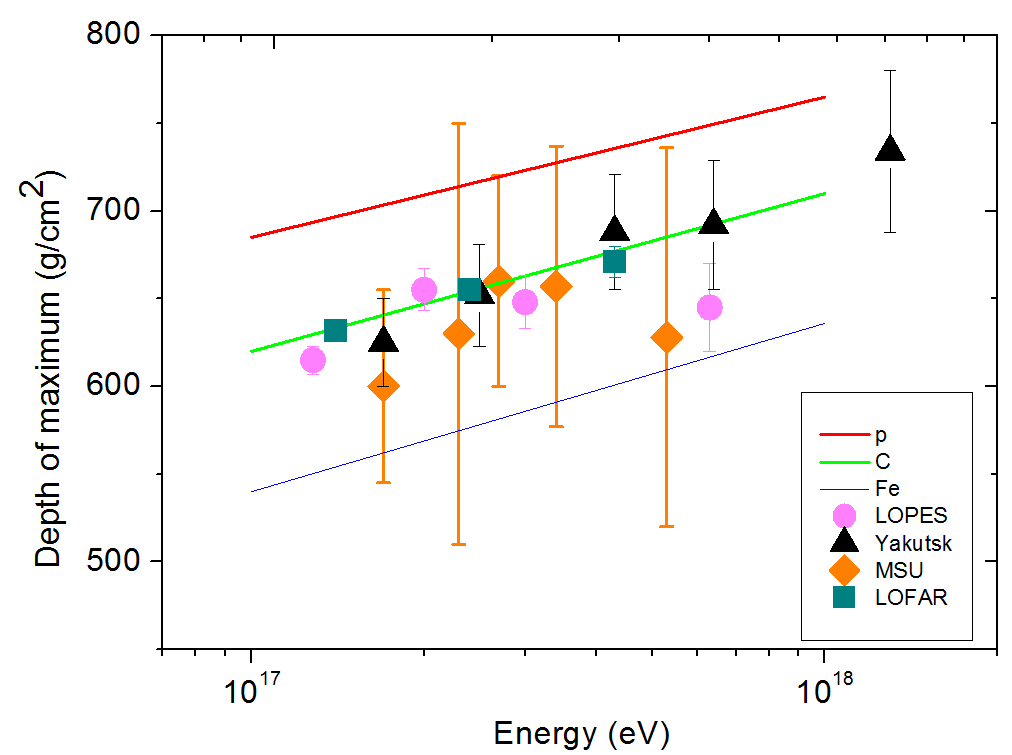}}
\caption{Comparison of experimental data of X$_{max}$, obtained at Yakutsk (triangles), MSU (diamonds)\cite{Kalmykov2013409}, LOPES (dots)\cite{Apel201285} and LOFAR \cite{Buitnik201490082003}}
\label{Fig:image17}
\end{figure}

Fig. ~\ref{Fig:image17} shows the dependence of X$_{max}$ from energy obtained by averaging a large number of individual showers. For comparison, results of LOPES, LOFAR and MSU (Moscow State University) are also plotted \cite{Kalmykov2013409, Apel201285, Buitnik201490082003}. Solid lines are model calculations of hadron interactions QGSjetII-04 for primary particles p, C, Fe \cite{Ostapchenko201183014018, Heck1988}. The experimental results within experimental error are consistent with each other. Still, large statistical errors do not allow us to make a definitive conclusion about the mass composition in the energy range 10$^{17}$ - 10$^{18}$ eV, but by comparison of obtained results with model calculations QGSjetII-04, the preliminary conclusion would be of the mixed composition of cosmic ray particles. In the future, with more statistics of air showers with radio measurements more specific conclusion about the mass composition can be made.

\section{Conclusion}
\label{yakutsk_sec_conclusion}

The Yakutsk array for more than 45 years continuously registers air showers in the energy range 10$^{15}$ - 10$^{20}$ eV. Constantly upgraded, extending the range of techniques to study the structure of spatial and longitudinal EAS development \cite{Artamonov1994929758, Knurenko2001157160}. Measurement of the radio emission from the EAS at the start of the Yakutsk experiment in 1970 was part of the research program and now we can conclude some results:
\begin{enumerate}
\item Long-term observation of air shower radio emission with ultra-high energies at the Yakutsk array shown that such measurements are promising at energies E$_0\geq$ 10$^{19}$ eV. In addition, a function of air shower radio emission attenuation at energies E$_0\geq$ 10$^{19}$ eV was obtained.
\item Empirically evaluated the shower energy E$_{0}$ from radio emission amplitude and X$_{max}$ from LDF of radio emission \cite{Knurenko2015410412}.
\item First shower with energy $\sim$2$\cdot$10$^{20}$ eV registered by antennas at Yakutsk array.
\item It was shown by measurement at Yakutsk array that the radio emission from the relativistic shower particles at energies above 10$^{17}$ eV can be used to study the physics of the air showers as an independent method and in conjunction with other methods of registering EAS for their mutual calibration \cite{Knurenko2015410412}.
\item In the future, the method of measuring radio emission, one can study the spectrum and mass composition of cosmic rays at energies above 10$^{19}$ eV not only registering radio emission on the surface \cite{Petrov2011510, Huege62012016} but also significantly increasing the aperture ratio of the array by placing the antenna on the space station, orbiting Earth \cite{Tsarev2004149}.

\end{enumerate}

\section*{Acknowledgments}

The reported study was funded by RFBR according to the research project 16-29-13019.

We wish to thank all of the staff of the Yakutsk Array for the opportunity to use experimental data and useful discussion during the article preparation.

\section*{References}

\bibliography{mybibfile}

\begin{thebibliography}{49}
\expandafter\ifx\csname natexlab\endcsname\relax\def\natexlab#1{#1}\fi
\expandafter\ifx\csname bibnamefont\endcsname\relax
  \def\bibnamefont#1{#1}\fi
\expandafter\ifx\csname bibfnamefont\endcsname\relax
  \def\bibfnamefont#1{#1}\fi
\expandafter\ifx\csname citenamefont\endcsname\relax
  \def\citenamefont#1{#1}\fi
\expandafter\ifx\csname url\endcsname\relax
  \def\url#1{\texttt{#1}}\fi
\expandafter\ifx\csname urlprefix\endcsname\relax\def\urlprefix{URL }\fi
\providecommand{\bibinfo}[2]{#2}
\providecommand{\eprint}[2][]{\url{#2}}

\bibitem[{\citenamefont{Kahn and Lerche}(1966)}]{Kahn2892061966}
\bibinfo{author}{\bibfnamefont{F.}~\bibnamefont{Kahn}} \bibnamefont{and}
  \bibinfo{author}{\bibfnamefont{I.}~\bibnamefont{Lerche}},
  \bibinfo{journal}{Proc. of R. Soc. A} \textbf{\bibinfo{volume}{289}},
  \bibinfo{pages}{206} (\bibinfo{year}{1966}).

\bibitem[{\citenamefont{Falcke and Gorham}(2003)}]{Falcke194772003}
\bibinfo{author}{\bibfnamefont{H.}~\bibnamefont{Falcke}} \bibnamefont{and}
  \bibinfo{author}{\bibfnamefont{P.}~\bibnamefont{Gorham}},
  \bibinfo{journal}{Astroparticle Physics} \textbf{\bibinfo{volume}{19}},
  \bibinfo{pages}{477} (\bibinfo{year}{2003}).

\bibitem[{\citenamefont{Askaryan}(1961)}]{Askaryan1961616}
\bibinfo{author}{\bibfnamefont{G.}~\bibnamefont{Askaryan}},
  \bibinfo{journal}{Sov. Phys. JETP} \textbf{\bibinfo{volume}{41}},
  \bibinfo{pages}{616} (\bibinfo{year}{1961}).

\bibitem[{\citenamefont{Jelley et~al.}(1965)\citenamefont{Jelley, Fruin,
  Porter, Weekes, Smith, and Porter}}]{Jelley1965327}
\bibinfo{author}{\bibfnamefont{J.}~\bibnamefont{Jelley}},
  \bibinfo{author}{\bibfnamefont{J.}~\bibnamefont{Fruin}},
  \bibinfo{author}{\bibfnamefont{N.}~\bibnamefont{Porter}},
  \bibinfo{author}{\bibfnamefont{T.}~\bibnamefont{Weekes}},
  \bibinfo{author}{\bibfnamefont{F.}~\bibnamefont{Smith}}, \bibnamefont{and}
  \bibinfo{author}{\bibfnamefont{R.}~\bibnamefont{Porter}},
  \bibinfo{journal}{Nature} \textbf{\bibinfo{volume}{205}},
  \bibinfo{pages}{327} (\bibinfo{year}{1965}).

\bibitem[{\citenamefont{Allan}(1971)}]{Allan1971171}
\bibinfo{author}{\bibfnamefont{H.}~\bibnamefont{Allan}},
  \bibinfo{journal}{Prog. in Elemen. Part. And Cos. Ray Phys}
  \textbf{\bibinfo{volume}{10}}, \bibinfo{pages}{171} (\bibinfo{year}{1971}).

\bibitem[{\citenamefont{Artamonov et~al.}(1987)\citenamefont{Artamonov,
  Atrashkevich, Egorov, and et~al.}}]{Artamonov1987109}
\bibinfo{author}{\bibfnamefont{V.}~\bibnamefont{Artamonov}},
  \bibinfo{author}{\bibfnamefont{V.}~\bibnamefont{Atrashkevich}},
  \bibinfo{author}{\bibfnamefont{T.}~\bibnamefont{Egorov}}, \bibnamefont{and}
  \bibinfo{author}{\bibnamefont{et~al.}}, \bibinfo{journal}{Proc. of Extensive
  Air Showers with Energy more than 10$^{17}$ eV} pp. \bibinfo{pages}{109--113}
  (\bibinfo{year}{1987}).

\bibitem[{\citenamefont{Tsarev}(2004)}]{Tsarev2004149}
\bibinfo{author}{\bibfnamefont{A.}~\bibnamefont{Tsarev}},
  \bibinfo{journal}{Physics of Elementary Particles and Atomic Nuclei}
  \textbf{\bibinfo{volume}{35}}, \bibinfo{pages}{1} (\bibinfo{year}{2004}).

\bibitem[{\citenamefont{Filonenko}(2015)}]{Filonenko2001439}
\bibinfo{author}{\bibfnamefont{A.}~\bibnamefont{Filonenko}},
  \bibinfo{journal}{Physics-Uspekhi} \textbf{\bibinfo{volume}{58}},
  \bibinfo{pages}{633} (\bibinfo{year}{2015}).

\bibitem[{\citenamefont{Huege}(2016)}]{Huege62012016}
\bibinfo{author}{\bibfnamefont{T.}~\bibnamefont{Huege}},
  \bibinfo{journal}{Physics reports} \textbf{\bibinfo{volume}{620}},
  \bibinfo{pages}{1} (\bibinfo{year}{2016}).

\bibitem[{\citenamefont{Schr{\"{o}}der}(2017)}]{Schroder931682017}
\bibinfo{author}{\bibfnamefont{F.}~\bibnamefont{Schr{\"{o}}der}},
  \bibinfo{journal}{Progress in Particle and Nuclear Physics}
  \textbf{\bibinfo{volume}{93}}, \bibinfo{pages}{1} (\bibinfo{year}{2017}).

\bibitem[{\citenamefont{Vernov et~al.}(1966)\citenamefont{Vernov, Abrosimov,
  Volovik, Zalyubovsky, and Christiansen}}]{Vernov1966157}
\bibinfo{author}{\bibfnamefont{S.}~\bibnamefont{Vernov}},
  \bibinfo{author}{\bibfnamefont{A.}~\bibnamefont{Abrosimov}},
  \bibinfo{author}{\bibfnamefont{V.}~\bibnamefont{Volovik}},
  \bibinfo{author}{\bibfnamefont{I.}~\bibnamefont{Zalyubovsky}},
  \bibnamefont{and}
  \bibinfo{author}{\bibfnamefont{G.}~\bibnamefont{Christiansen}},
  \bibinfo{journal}{JETP Letters} pp. \bibinfo{pages}{157--162}
  (\bibinfo{year}{1966}).

\bibitem[{\citenamefont{Atrashkevich et~al.}(1978)\citenamefont{Atrashkevich,
  Vedeneev, Alan, and et~al.}}]{Atrashkevich1978712}
\bibinfo{author}{\bibfnamefont{V.}~\bibnamefont{Atrashkevich}},
  \bibinfo{author}{\bibfnamefont{O.}~\bibnamefont{Vedeneev}},
  \bibinfo{author}{\bibfnamefont{C.}~\bibnamefont{Alan}}, \bibnamefont{and}
  \bibinfo{author}{\bibnamefont{et~al.}}, \bibinfo{journal}{Nuclear Physics}
  \textbf{\bibinfo{volume}{28}}, \bibinfo{pages}{712} (\bibinfo{year}{1978}).

\bibitem[{\citenamefont{Artamonov et~al.}(1988)\citenamefont{Artamonov,
  Atrashkevich, Egorov, and et~al.}}]{Artamonov19884748}
\bibinfo{author}{\bibfnamefont{V.}~\bibnamefont{Artamonov}},
  \bibinfo{author}{\bibfnamefont{V.}~\bibnamefont{Atrashkevich}},
  \bibinfo{author}{\bibfnamefont{T.}~\bibnamefont{Egorov}}, \bibnamefont{and}
  \bibinfo{author}{\bibnamefont{et~al.}}, \bibinfo{journal}{Proc. of All-Union
  conference of cosmic rays} \textbf{\bibinfo{volume}{1}}, \bibinfo{pages}{47}
  (\bibinfo{year}{1988}).

\bibitem[{\citenamefont{Artamonov et~al.}(1990)\citenamefont{Artamonov, Egorov,
  Efimov, and et~al.}}]{Artamonov1990210}
\bibinfo{author}{\bibfnamefont{V.}~\bibnamefont{Artamonov}},
  \bibinfo{author}{\bibfnamefont{T.}~\bibnamefont{Egorov}},
  \bibinfo{author}{\bibfnamefont{A.~N.} \bibnamefont{Efimov}},
  \bibnamefont{and} \bibinfo{author}{\bibnamefont{et~al.}},
  \bibinfo{journal}{Proc. 21st ICRC} \textbf{\bibinfo{volume}{9}},
  \bibinfo{pages}{210} (\bibinfo{year}{1990}).

\bibitem[{\citenamefont{Schellart et~al.}(2013)\citenamefont{Schellart, Nelles,
  Buitink, and et~al.}}]{Schellart2013560}
\bibinfo{author}{\bibfnamefont{P.}~\bibnamefont{Schellart}},
  \bibinfo{author}{\bibfnamefont{A.}~\bibnamefont{Nelles}},
  \bibinfo{author}{\bibfnamefont{S.}~\bibnamefont{Buitink}}, \bibnamefont{and}
  \bibinfo{author}{\bibnamefont{et~al.}}, \bibinfo{journal}{A$\&$A}
  \textbf{\bibinfo{volume}{560}}, \bibinfo{pages}{A98} (\bibinfo{year}{2013}).

\bibitem[{\citenamefont{Fuchs and et~al.}(2012)}]{Fuchs201293}
\bibinfo{author}{\bibfnamefont{B.}~\bibnamefont{Fuchs}} \bibnamefont{and}
  \bibinfo{author}{\bibnamefont{et~al.}}, \bibinfo{journal}{Nucl. Instr. and
  Meth. A} \textbf{\bibinfo{volume}{692}}, \bibinfo{pages}{93}
  (\bibinfo{year}{2012}).

\bibitem[{\citenamefont{Apel and et~al.}(2014)}]{Apel201490}
\bibinfo{author}{\bibfnamefont{W.}~\bibnamefont{Apel}} \bibnamefont{and}
  \bibinfo{author}{\bibnamefont{et~al.}}, \bibinfo{journal}{Physical Review D}
  \textbf{\bibinfo{volume}{90}}, \bibinfo{pages}{062001}
  (\bibinfo{year}{2014}).

\bibitem[{\citenamefont{Knurenko and
  Petrov}(2015{\natexlab{a}})}]{Knurenko2015632}
\bibinfo{author}{\bibfnamefont{S.}~\bibnamefont{Knurenko}} \bibnamefont{and}
  \bibinfo{author}{\bibfnamefont{I.}~\bibnamefont{Petrov}},
  \bibinfo{journal}{J. Phys.: Conf. Ser.} \textbf{\bibinfo{volume}{632}},
  \bibinfo{pages}{012100} (\bibinfo{year}{2015}{\natexlab{a}}).

\bibitem[{\citenamefont{Matthews}(2013)}]{Matthews20131218}
\bibinfo{author}{\bibfnamefont{J.}~\bibnamefont{Matthews}},
  \bibinfo{journal}{Proc. of 33rd ICRC} p. \bibinfo{pages}{1218}
  (\bibinfo{year}{2013}).

\bibitem[{\citenamefont{Bezyazeekov et~al.}(2016)\citenamefont{Bezyazeekov,
  Budnev, Gress, and et~al.}}]{Schroder010522016}
\bibinfo{author}{\bibfnamefont{P.}~\bibnamefont{Bezyazeekov}},
  \bibinfo{author}{\bibfnamefont{N.}~\bibnamefont{Budnev}},
  \bibinfo{author}{\bibfnamefont{O.}~\bibnamefont{Gress}}, \bibnamefont{and}
  \bibinfo{author}{\bibnamefont{et~al.}}, \bibinfo{journal}{JCAP}
  \textbf{\bibinfo{volume}{01}}, \bibinfo{pages}{052} (\bibinfo{year}{2016}).

\bibitem[{\citenamefont{Apel et~al.}(2016)\citenamefont{Apel,
  {Arteaga-Vel{\'{a}}squez}, Bahren, and et~al.}}]{Hiller7631792016}
\bibinfo{author}{\bibfnamefont{W.}~\bibnamefont{Apel}},
  \bibinfo{author}{\bibfnamefont{J.}~\bibnamefont{{Arteaga-Vel{\'{a}}squez}}},
  \bibinfo{author}{\bibfnamefont{L.}~\bibnamefont{Bahren}}, \bibnamefont{and}
  \bibinfo{author}{\bibnamefont{et~al.}}, \bibinfo{journal}{Physics Letters B}
  \textbf{\bibinfo{volume}{763}}, \bibinfo{pages}{179} (\bibinfo{year}{2016}).

\bibitem[{\citenamefont{Aab and et~al.}(2016)}]{Aab93120052016}
\bibinfo{author}{\bibfnamefont{A.}~\bibnamefont{Aab}} \bibnamefont{and}
  \bibinfo{author}{\bibnamefont{et~al.}}, \bibinfo{journal}{Phys. Rev.D}
  \textbf{\bibinfo{volume}{93}}, \bibinfo{pages}{12005} (\bibinfo{year}{2016}).

\bibitem[{\citenamefont{Kozlov et~al.}(2012)\citenamefont{Kozlov, Knurenko,
  Mullayarov, Petrov, and Pravdin}}]{Kozlov2012215}
\bibinfo{author}{\bibfnamefont{V.}~\bibnamefont{Kozlov}},
  \bibinfo{author}{\bibfnamefont{S.}~\bibnamefont{Knurenko}},
  \bibinfo{author}{\bibfnamefont{V.}~\bibnamefont{Mullayarov}},
  \bibinfo{author}{\bibfnamefont{Z.}~\bibnamefont{Petrov}}, \bibnamefont{and}
  \bibinfo{author}{\bibfnamefont{M.}~\bibnamefont{Pravdin}},
  \bibinfo{journal}{Proc. 1st Int. Conf. Electromagnetic Method of
  Environmental Studies} pp. \bibinfo{pages}{215--–217}
  (\bibinfo{year}{2012}).

\bibitem[{\citenamefont{Ellingson et~al.}(2007)\citenamefont{Ellingson,
  Simonetti, and Patterson}}]{Ellingson5532007}
\bibinfo{author}{\bibfnamefont{S.}~\bibnamefont{Ellingson}},
  \bibinfo{author}{\bibfnamefont{J.}~\bibnamefont{Simonetti}},
  \bibnamefont{and}
  \bibinfo{author}{\bibfnamefont{C.}~\bibnamefont{Patterson}},
  \bibinfo{journal}{IEEE Trans. Antennas and Propag.}
  \textbf{\bibinfo{volume}{55}}, \bibinfo{pages}{826} (\bibinfo{year}{2007}).

\bibitem[{\citenamefont{Petrov et~al.}(2011)\citenamefont{Petrov, Borschevsky,
  Knurenko, Kozlov, and Petrov}}]{Petrov2011510}
\bibinfo{author}{\bibfnamefont{Z.}~\bibnamefont{Petrov}},
  \bibinfo{author}{\bibfnamefont{D.}~\bibnamefont{Borschevsky}},
  \bibinfo{author}{\bibfnamefont{S.}~\bibnamefont{Knurenko}},
  \bibinfo{author}{\bibfnamefont{V.}~\bibnamefont{Kozlov}}, \bibnamefont{and}
  \bibinfo{author}{\bibfnamefont{I.}~\bibnamefont{Petrov}},
  \bibinfo{journal}{NEFU Herald} \textbf{\bibinfo{volume}{8}},
  \bibinfo{pages}{5} (\bibinfo{year}{2011}).

\bibitem[{\citenamefont{Knurenko and Petrov}(2016)}]{Knurenko20161045297}
\bibinfo{author}{\bibfnamefont{S.}~\bibnamefont{Knurenko}} \bibnamefont{and}
  \bibinfo{author}{\bibfnamefont{I.}~\bibnamefont{Petrov}},
  \bibinfo{journal}{JETP Lett.} \textbf{\bibinfo{volume}{104}},
  \bibinfo{pages}{297} (\bibinfo{year}{2016}).

\bibitem[{\citenamefont{Huege et~al.}(2008)\citenamefont{Huege, {Ulrich}, and
  Engel}}]{Huege200830}
\bibinfo{author}{\bibfnamefont{T.}~\bibnamefont{Huege}},
  \bibinfo{author}{\bibfnamefont{R.}~\bibnamefont{{Ulrich}}}, \bibnamefont{and}
  \bibinfo{author}{\bibfnamefont{R.}~\bibnamefont{Engel}},
  \bibinfo{journal}{Astropart. Phys.} \textbf{\bibinfo{volume}{30}},
  \bibinfo{pages}{96} (\bibinfo{year}{2008}).

\bibitem[{\citenamefont{Kalmykov et~al.}(2013)\citenamefont{Kalmykov,
  Konstantinov, and Vedeneev}}]{Kalmykov2013409}
\bibinfo{author}{\bibfnamefont{N.}~\bibnamefont{Kalmykov}},
  \bibinfo{author}{\bibfnamefont{A.}~\bibnamefont{Konstantinov}},
  \bibnamefont{and} \bibinfo{author}{\bibfnamefont{O.}~\bibnamefont{Vedeneev}},
  \bibinfo{journal}{Journal of Physics: Conference series}
  \textbf{\bibinfo{volume}{409}}, \bibinfo{pages}{2012071}
  (\bibinfo{year}{2013}).

\bibitem[{\citenamefont{Apel et~al.}(2012)\citenamefont{Apel, Arteaga, and
  B{\"{a}}hren}}]{Apel201285}
\bibinfo{author}{\bibfnamefont{W.}~\bibnamefont{Apel}},
  \bibinfo{author}{\bibfnamefont{J.}~\bibnamefont{Arteaga}}, \bibnamefont{and}
  \bibinfo{author}{\bibfnamefont{L.}~\bibnamefont{B{\"{a}}hren}},
  \bibinfo{journal}{Physical Review D} \textbf{\bibinfo{volume}{85}},
  \bibinfo{pages}{071101(R)} (\bibinfo{year}{2012}).

\bibitem[{\citenamefont{Knurenko and
  Petrov}(2015{\natexlab{b}})}]{Knurenko2015410412}
\bibinfo{author}{\bibfnamefont{S.}~\bibnamefont{Knurenko}} \bibnamefont{and}
  \bibinfo{author}{\bibfnamefont{I.}~\bibnamefont{Petrov}},
  \bibinfo{journal}{Bulletin of the Russian Academy of Science. Physics}
  \textbf{\bibinfo{volume}{79}}, \bibinfo{pages}{410}
  (\bibinfo{year}{2015}{\natexlab{b}}).

\bibitem[{\citenamefont{Knurenko and Petrov}(2013)}]{Knurenko20130054}
\bibinfo{author}{\bibfnamefont{S.}~\bibnamefont{Knurenko}} \bibnamefont{and}
  \bibinfo{author}{\bibfnamefont{I.}~\bibnamefont{Petrov}},
  \bibinfo{journal}{Proceedings of the 33th International Cosmic Ray
  Conference} p. \bibinfo{pages}{0054} (\bibinfo{year}{2013}).

\bibitem[{\citenamefont{Schr{\"{o}}der and et~al}(2012)}]{Schroder2012662}
\bibinfo{author}{\bibfnamefont{F.}~\bibnamefont{Schr{\"{o}}der}}
  \bibnamefont{and} \bibinfo{author}{\bibnamefont{et~al}},
  \bibinfo{journal}{Nucl. Instr. and Meth. A} \textbf{\bibinfo{volume}{662}},
  \bibinfo{pages}{S238} (\bibinfo{year}{2012}).

\bibitem[{\citenamefont{Dyakonov et~al.}(1991)\citenamefont{Dyakonov, Egorov,
  Efimov, and et~al.}}]{Dyakonov2521991}
\bibinfo{author}{\bibfnamefont{M.}~\bibnamefont{Dyakonov}},
  \bibinfo{author}{\bibfnamefont{T.}~\bibnamefont{Egorov}},
  \bibinfo{author}{\bibfnamefont{N.}~\bibnamefont{Efimov}}, \bibnamefont{and}
  \bibinfo{author}{\bibnamefont{et~al.}}, \bibinfo{journal}{Science,
  Novosibirsk} p. \bibinfo{pages}{252} (\bibinfo{year}{1991}).

\bibitem[{\citenamefont{Knurenko et~al.}(2006)\citenamefont{Knurenko, Ivanov,
  Sleptsov, and Sabourov}}]{Knurenko83114732006}
\bibinfo{author}{\bibfnamefont{S.}~\bibnamefont{Knurenko}},
  \bibinfo{author}{\bibfnamefont{A.}~\bibnamefont{Ivanov}},
  \bibinfo{author}{\bibfnamefont{I.}~\bibnamefont{Sleptsov}}, \bibnamefont{and}
  \bibinfo{author}{\bibfnamefont{A.}~\bibnamefont{Sabourov}},
  \bibinfo{journal}{JETP Lett.} \textbf{\bibinfo{volume}{83}},
  \bibinfo{pages}{473} (\bibinfo{year}{2006}).

\bibitem[{\citenamefont{Ivanov et~al.}(2007)\citenamefont{Ivanov, Knurenko, and
  Sleptsov}}]{Ivanov131610012007}
\bibinfo{author}{\bibfnamefont{A.}~\bibnamefont{Ivanov}},
  \bibinfo{author}{\bibfnamefont{S.}~\bibnamefont{Knurenko}}, \bibnamefont{and}
  \bibinfo{author}{\bibfnamefont{I.}~\bibnamefont{Sleptsov}},
  \bibinfo{journal}{JETP} \textbf{\bibinfo{volume}{131}}, \bibinfo{pages}{1001}
  (\bibinfo{year}{2007}).

\bibitem[{\citenamefont{Glushkov et~al.}(1953)\citenamefont{Glushkov,
  M.N.~Dyakonov, and et~al.}}]{Glushkov57911993}
\bibinfo{author}{\bibfnamefont{A.}~\bibnamefont{Glushkov}},
  \bibinfo{author}{\bibfnamefont{T.~E.} \bibnamefont{M.N.~Dyakonov}},
  \bibnamefont{and} \bibinfo{author}{\bibnamefont{et~al.}},
  \bibinfo{journal}{Bulletin of the Russian Academy of Science. Physics}
  \textbf{\bibinfo{volume}{57}}, \bibinfo{pages}{91} (\bibinfo{year}{1953}).

\bibitem[{\citenamefont{Knurenko et~al.}(2013)\citenamefont{Knurenko, Kozlov,
  Petrov, and Pravdin}}]{Knurenko771115592013}
\bibinfo{author}{\bibfnamefont{S.}~\bibnamefont{Knurenko}},
  \bibinfo{author}{\bibfnamefont{V.}~\bibnamefont{Kozlov}},
  \bibinfo{author}{\bibfnamefont{Z.}~\bibnamefont{Petrov}}, \bibnamefont{and}
  \bibinfo{author}{\bibfnamefont{M.}~\bibnamefont{Pravdin}},
  \bibinfo{journal}{Bulletin of Russian Academy of Sciences. Physics}
  \textbf{\bibinfo{volume}{77}}, \bibinfo{pages}{1559} (\bibinfo{year}{2013}).

\bibitem[{\citenamefont{Horneffer et~al.}(2007)\citenamefont{Horneffer, Apel,
  Arteaga, and et~al.}}]{Horneffer20088386}
\bibinfo{author}{\bibfnamefont{A.}~\bibnamefont{Horneffer}},
  \bibinfo{author}{\bibfnamefont{W.}~\bibnamefont{Apel}},
  \bibinfo{author}{\bibfnamefont{J.}~\bibnamefont{Arteaga}}, \bibnamefont{and}
  \bibinfo{author}{\bibnamefont{et~al.}}, \bibinfo{journal}{Proc. of the 30th
  ICRC, Merida, Mexico} \textbf{\bibinfo{volume}{4}}, \bibinfo{pages}{83}
  (\bibinfo{year}{2007}).

\bibitem[{\citenamefont{Apel and et~al.}(2010)}]{Apel2010202216620}
\bibinfo{author}{\bibfnamefont{W.}~\bibnamefont{Apel}} \bibnamefont{and}
  \bibinfo{author}{\bibnamefont{et~al.}}, \bibinfo{journal}{Nucl. Instr. and
  Meth. A} \textbf{\bibinfo{volume}{620}}, \bibinfo{pages}{202}
  (\bibinfo{year}{2010}).

\bibitem[{\citenamefont{Tikhonov and Arsenin}(1977)}]{Tikhonov2581977}
\bibinfo{author}{\bibfnamefont{A.}~\bibnamefont{Tikhonov}} \bibnamefont{and}
  \bibinfo{author}{\bibfnamefont{V.}~\bibnamefont{Arsenin}},
  \bibinfo{journal}{Winston, NY} p. \bibinfo{pages}{258}
  (\bibinfo{year}{1977}).

\bibitem[{\citenamefont{Dyakonov et~al.}(1988)\citenamefont{Dyakonov, Knurenko,
  Kolosov, and et~al.}}]{Dyakonov1986224226}
\bibinfo{author}{\bibfnamefont{M.}~\bibnamefont{Dyakonov}},
  \bibinfo{author}{\bibfnamefont{S.}~\bibnamefont{Knurenko}},
  \bibinfo{author}{\bibfnamefont{V.}~\bibnamefont{Kolosov}}, \bibnamefont{and}
  \bibinfo{author}{\bibnamefont{et~al.}}, \bibinfo{journal}{Nucl. Instr. and
  Meth. A} \textbf{\bibinfo{volume}{248}}, \bibinfo{pages}{224}
  (\bibinfo{year}{1988}).

\bibitem[{\citenamefont{Knurenko et~al.}(2001)\citenamefont{Knurenko, Kolosov,
  Petrov, and et~al.}}]{Knurenko2001157160}
\bibinfo{author}{\bibfnamefont{S.}~\bibnamefont{Knurenko}},
  \bibinfo{author}{\bibfnamefont{V.}~\bibnamefont{Kolosov}},
  \bibinfo{author}{\bibfnamefont{Z.}~\bibnamefont{Petrov}}, \bibnamefont{and}
  \bibinfo{author}{\bibnamefont{et~al.}}, \bibinfo{journal}{Proc. of 27th ICRC}
  \textbf{\bibinfo{volume}{1}}, \bibinfo{pages}{157} (\bibinfo{year}{2001}).

\bibitem[{\citenamefont{Kochev}(1985)}]{Kochev62711985}
\bibinfo{author}{\bibfnamefont{V.}~\bibnamefont{Kochev}}, \bibinfo{journal}{Use
  of Computers in Control Problems} pp. \bibinfo{pages}{62--71}
  (\bibinfo{year}{1985}).

\bibitem[{\citenamefont{Dyakonov et~al.}(1999)\citenamefont{Dyakonov, Knurenko,
  Kolosov, and et~al.}}]{Dyakonov3153191999}
\bibinfo{author}{\bibfnamefont{M.}~\bibnamefont{Dyakonov}},
  \bibinfo{author}{\bibfnamefont{S.}~\bibnamefont{Knurenko}},
  \bibinfo{author}{\bibfnamefont{V.}~\bibnamefont{Kolosov}}, \bibnamefont{and}
  \bibinfo{author}{\bibnamefont{et~al.}}, \bibinfo{journal}{Atmos. Oceanic.
  Optic.} \textbf{\bibinfo{volume}{12}}, \bibinfo{pages}{315}
  (\bibinfo{year}{1999}).

\bibitem[{\citenamefont{Knurenko and Sabourov}(2011)}]{Knurenko72512552011}
\bibinfo{author}{\bibfnamefont{S.}~\bibnamefont{Knurenko}} \bibnamefont{and}
  \bibinfo{author}{\bibfnamefont{A.}~\bibnamefont{Sabourov}},
  \bibinfo{journal}{Astrophys. Space Sci. Trans.} \textbf{\bibinfo{volume}{7}},
  \bibinfo{pages}{251} (\bibinfo{year}{2011}).

\bibitem[{\citenamefont{Buitnik and et~al.}(2014)}]{Buitnik201490082003}
\bibinfo{author}{\bibfnamefont{S.}~\bibnamefont{Buitnik}} \bibnamefont{and}
  \bibinfo{author}{\bibnamefont{et~al.}}, \bibinfo{journal}{Physical Review D}
  \textbf{\bibinfo{volume}{90}}, \bibinfo{pages}{082003}
  (\bibinfo{year}{2014}).

\bibitem[{\citenamefont{Ostapchenko}(2011)}]{Ostapchenko201183014018}
\bibinfo{author}{\bibfnamefont{S.}~\bibnamefont{Ostapchenko}},
  \bibinfo{journal}{Physical Review D} \textbf{\bibinfo{volume}{83}},
  \bibinfo{pages}{014018} (\bibinfo{year}{2011}).

\bibitem[{\citenamefont{Heck et~al.}(1998)\citenamefont{Heck, Knapp,
  Capdevielle, and et~al.}}]{Heck1988}
\bibinfo{author}{\bibfnamefont{D.}~\bibnamefont{Heck}},
  \bibinfo{author}{\bibfnamefont{J.}~\bibnamefont{Knapp}},
  \bibinfo{author}{\bibfnamefont{J.}~\bibnamefont{Capdevielle}},
  \bibnamefont{and} \bibinfo{author}{\bibnamefont{et~al.}},
  \bibinfo{journal}{Report-FZKA, Forschungszentrum Karlsruhe}
  \textbf{\bibinfo{volume}{248}}, \bibinfo{pages}{6019} (\bibinfo{year}{1998}).

\bibitem[{\citenamefont{Artamonov and et~al.}(1994)}]{Artamonov1994929758}
\bibinfo{author}{\bibfnamefont{V.}~\bibnamefont{Artamonov}} \bibnamefont{and}
  \bibinfo{author}{\bibnamefont{et~al.}}, \bibinfo{journal}{Bulletin of Russian
  Academy of Science. Physics} \textbf{\bibinfo{volume}{58}},
  \bibinfo{pages}{92} (\bibinfo{year}{1994}).

\end{thebibliography}

\end{document}